\documentclass[usenatbib,usegraphicx]{mn2e}

\title[Current star formation in early-type galaxies]{Current star formation in early-type galaxies and the K+A phenomenon}
\author[J. Helmboldt et al.]{J. F. Helmboldt$^{1,2,3}$\thanks{E-mail: helmbold@unm.edu}, R. A. M. Walterbos$^{3}$ and T. Goto$^{4}$ \\
$^{1}$Naval Research Laboratory, Code 7213, 4555 Overlook Avenue SW, Washington, DC 20375-5351, USA \\
$^{2}$Department of Physics and Astronomy, University of New Mexico, 800 Yale Blvd NE, Albuquerque, NM 87131, USA \\
$^{3}$Department of Astronomy, New Mexico State University, MSC 4500 PO Box 30001 Las Cruces, NM 88003, USA \\
$^{4}$Institute of Space and Astronautical Science, Japan Aerospace Exploration Agency, Sagamihara, Kanagawa, 229-8510, Japan \\}
\begin{document}

\date{Not yet submitted.}

\pagerange{\pageref{firstpage}--\pageref{lastpage}} \pubyear{2007}

\maketitle

\label{firstpage}

\begin{abstract}
We present the results of an effort to identify and study a sample of the likely progenitors of elliptical (E) and lenticular (S0) K+A galaxies.  To achieve this, we have searched a sample $\sim$11,000 nearby ($m_{r}<$16) early-type galaxies selected by morphology from the Sloan Digital Sky Survey (SDSS) Main spectroscopic sample for actively star-forming E and S0 galaxies.  Using emission line ratios and visual inspection of SDSS $g$-band images, we have identified 335 galaxies from the SDSS Fourth Data Release (DR4) as actively star-forming E and S0 galaxies.  These galaxies make up about 3\% of the total early-type sample and less than 1\% of all Main galaxies with $m_{r}<$16.  We also identified a sample of $\sim$400 K+A galaxies from DR4 with $m_{r}<$16; more than half of these are E and S0 galaxies.  We find that star-forming early-type galaxies and K+A galaxies have similar mass distributions; they are on average less massive than typical early-type galaxies but more massive than the average star-forming galaxy.  Both of these types of galaxies are found in higher fractions among all galaxies in lower density environments.  The fractions of star-forming E and S0 galaxies and E and S0 K+A galaxies depend on environment in nearly the same way.  Model spectra fit to the stellar continua of the star-forming E and S0 galaxies showed that their properties are consistent with star formation episodes of $\;^{<}_{\sim}$1 Gyr in duration.  The modelling results imply that on average, the star formation episodes will increase the stellar masses by about 4\%.  The results also imply that only episodes that increase the stellar mass by more than 2-5\% will lead to K+A galaxies as we have defined them and that this is true for roughly 80\% of the star forming E and S0 galaxies in our sample.  The estimated typical increase in stellar mass implies that new stellar components of about $2\times 10^{8}$ M$_{\odot}$ will be formed on average.  There is also evidence that the star-forming regions within these galaxies are rotationally supported.  These two results, when combined with the galaxies' total masses and lack of prominent disks, suggest that the star formation within these galaxies may be confined to relatively small, central disks, similar to the nuclear stellar and dust disks found in many low mass early-type galaxies.

\end{abstract}

\begin{keywords}
galaxies: elliptical and lenticular, cD -- galaxies: star-burst -- galaxies: stellar content
\end{keywords}

\section{Introduction}

The formation and evolution of early-type (elliptical and lenticular) galaxies is crucial to understanding galaxy formation.  The fact that the stellar populations in typical early-type galaxies seem to have formed long ago \citep[$\sim$10 Gyr; e.g.,][]{bow92,ter01,tho05} appears to support the hypothesis that these galaxies formed relatively quickly at high redshift and have had little or no new star formation since then.  However, there is evidence that the stellar populations of early-type field galaxies may have formed at lower redshifts \citep{van01,tre02}.  There are even some early-type galaxies that show evidence of star formation within the last $\sim$1 Gyr \citep{zab96,got03,qui04,yi05,kav07}.  These galaxies belong to a larger class of galaxies referred to as K+A (or E+A\footnote{For this paper, we have chosen to use K+A rather than E+A since these galaxies are classified purely by their spectra, and the E+A designation implies some degree of morphological classification.}) galaxies discovered by \citet{dre83}.  K+A galaxies are defined by their relatively strong stellar Balmer absorption lines and lack of nebular emission lines (usually H$\alpha$ and/or $[$OII$]$).  These characteristics imply that while these galaxies formed some of their stars within the last $\sim$1 Gyr, they are not currently forming stars.  Recent studies of relatively large samples of K+A galaxies \citep{qui04,got03} have revealed that these galaxies are preferentially found outside of galaxy clusters, implying that they may provide useful information regarding the evolution of early-type field galaxies.  A detailed study of the environments of 266 K+A galaxies performed by \citet{got05} has shown that the local galaxy density within 100 kpc of these galaxies is larger than what is found for other galaxies; no such excess was found on size scales typical of galaxy clusters or large scale structure.  This implies that these galaxies may also be tracers of galaxy-galaxy tidal interactions or possibly merger/accretion events in the field.\par
Understanding the role K+A galaxies play within the domain of galaxy evolution depends highly on our knowledge of the mechanisms which trigger and halt the star formation episodes that lead to these objects.  To constrain the possible variety of such mechanisms and to obtain a more complete picture of these objects, a sample of likely progenitors of K+A galaxies is required.  While most K+A galaxies are early-type galaxies, many also have prominent disk components and spiral structure \citep{got05}.  Identifying the progenitors of these K+A spiral galaxies may be difficult since current star formation is commonplace among spiral galaxies.  A K+A spiral may be the result of the cessation of star formation within a relatively normal spiral as its gas is stripped via tidal interactions or ram pressure stripping within a group or cluster of galaxies \citep[e.g.,][]{ver01,vog04}.  A K+A spiral may also be produced by a relatively short episode of star formation contained within the nucleus of a bulge dominated spiral.  Conversely, current star formation is quite rare among early-type galaxies, and it is extremely unlikely that a process such as gas stripping could produce an early-type K+A galaxy.  While some have argued that mergers of spiral galaxies could produce early-type K+A galaxies \citep{bek05}, the optical and near-IR properties of the majority of K+A galaxies are consistent with the K+A spectral signature of these galaxies resulting from a short episode of star formation superimposed on an older stellar population \citep{bal05}.  It therefore seems promising to try to cull a sample of candidate K+A progenitors by identifying early-type galaxies that are actively forming stars.\par
Recently, \citet{fuk04} have discovered two galaxies in the Sloan Digital Sky Survey (SDSS) that are unambiguously elliptical galaxies that are currently forming stars.  To investigate the possibility that star-forming early-type galaxies like the elliptical galaxies discovered by \citet{fuk04} are plausibly the progenitors of early-type K+A galaxies, we have selected a large sample of actively star-forming early-type galaxies using the SDSS spectroscopic data and images.  The sample selection, including the definition of different galaxy categories, is detailed in \S 2.  The general properties of the sample galaxies and how they compare to K+A galaxies and early-type galaxies in general is discussed in \S 3.  The distribution of star formation time scales among star forming early-type galaxies is explored using population synthesis models.  The model implementation and results are discussed in \S 4; further applications of the model results are discussed in \S 5.  In \S 6, the main results are summarised and useful follow-up observations of this sample are discussed.\par

\section{Sample selection and galaxy categories}
\subsection{Identifying star-forming galaxies}
The selection of our sample of actively star-forming early-type galaxies consists of two basic parts.  The first involves the selection of a large sample of actively star-forming galaxies.  The second consists of the combined use of morphological selection criteria and visual inspection of optical images  to produce from this sample a sub-sample dominated by elliptical and lenticular galaxies.  We stress that the selection process was designed to yield a relatively clean sample of star forming elliptical and lenticular galaxies and is somewhat complex.  This makes a meaningful comparison of the number of galaxies in the resulting sample to those in other samples difficult, and such comparisons are not part of the analysis and interpretation that follows.  The two parts of this selection process are detailed below.\par
All of the galaxy data used here are taken from the fourth data release (DR4) of the SDSS \citep{ade06}.  The SDSS is an imaging an spectroscopic survey of 8,000 square degrees of the northern sky.  From $u \: g \: r \: i \: z$ images, $\sim 10^{6}$ spectroscopic targets were selected and observed with a fibre fed spectrograph that can observe 640 objects simultaneously \citep[see][]{gun98,yor00,sto02}.  These targets were selected in an automated way \citep{eis01,str02,ric02} based on photometric properties measured within an automated data processing pipeline that performs astrometry \citep{pie03}, source identification, de-blending, photometry \citep{lup01}, and calibration \citep{fuk96,hog01,smi02,ive04}.  The fibre placement was also automatically determined for the targeted objects \citep{bla03a}.\par
We have identified star-forming galaxies within the flux limited Main sample \citep{str02} from DR4.  While the Main sample has a flux limit of $m_{r}<17.77$, we have used a limiting magnitude of $m_{r}<16$ for reasons explained below.  To identify actively star forming galaxies, we have used emission line ratios and a diagnostic diagram similar to those used by \citet{ost06}.  For this purpose, we have used the emission line fluxes measured from SDSS spectra by \citet{tre04}.  The basic procedure used by \cite{tre04} is as follows.  The fluxes were measured by first fitting to each galaxy spectrum a linear combination of model spectra with different star formation histories produced using the population synthesis code of \citet{bru03}.  This fit was then subtracted to remove the stellar continuum, a step which is vital to our efforts to obtain accurate emission line fluxes for early-type galaxies whose stellar continua, especially near the galaxies' centres, are typically quite bright.  After subtraction of the model continuum fits, Gaussian functions were fit to the galaxy's emission lines.  From these fits, the flux, equivalent width, and velocity width was obtained for each emission line as well as estimates for the errors in each of these quantities.\par
To complete the first step in our selection process, we have chosen to use the emission line ratios $[$OIII$]$/H$\beta$ and $[$NII$]$/H$\alpha$ to select actively star-forming galaxies and exclude Seyfert and LINER galaxies.  We have chosen to only include galaxies with $m_{r}<$16 that are nearby enough that any spiral arms or prominent disks will be apparent in their $g$-band images.  We have chosen this flux limit because it similar to the limit used by \citet{nak03} for their catalogue of morphologically classified SDSS galaxies.  They have demonstrated that from the SDSS images, it is possible to obtain reasonably accurate morphological classifications by eye for galaxies down to this magnitude limit.  Formally, the limiting $r^{\prime}$-band magnitude used by \citet{nak03} was 15.9.  However, the fluxes they used were obtained using an older version of the SDSS photometric pipeline using the photometric system of SDSS early data release \citep[EDR; ][]{sto02}, the $u^{\prime},g^{\prime},r^{\prime},i^{\prime},z^{\prime}$ system.  In the photometric system used in subsequent data releases (i.e., $u,g,r,i,z$), including DR4, 4-5\% of the galaxies from their sample have $m_{r}>$15.9.  In light of this, we have chosen to increase the limiting magnitude to 16.\par
\begin{figure*}
\includegraphics[scale=0.75]{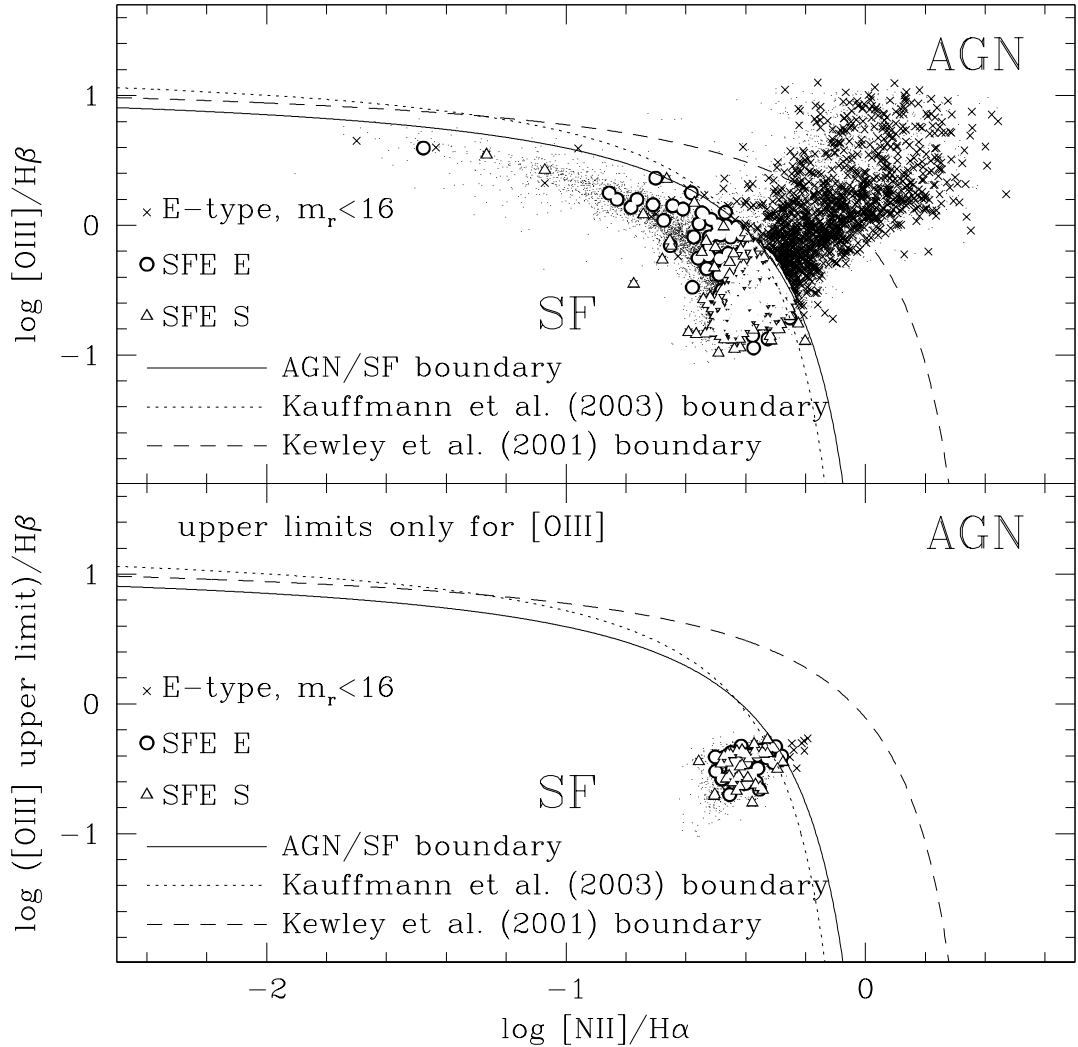}
\caption{For galaxies with $m_{r}<$16 and $\geq 5 \sigma$ detections of the $[$NII$] \lambda$6583, H$\beta$, and H$\alpha$ emission lines, $[$OIII$]$/H$\beta$ versus $[$NII$]$/H$\alpha$ for galaxies with $\geq 3 \sigma$ detections of the $[$OIII$] \lambda$5007 line (upper panel) and upper limits for $[$OIII$]$/H$\beta$ versus $[$NII$]$/H$\alpha$ for galaxies with no significant detection of $[$OIII$]$ emission (lower panel).  In both panels, the dashed curve is the upper limit for star-burst galaxies from \citet{kew01}, the dotted line is the empirical upper limit for star forming galaxies used by \citet{kau03a}, and the solid curve is the more conservative limit used here.  Early-type galaxies (see \S 2) are represented by $\times$'s; star forming early-type (SFE) galaxies are represented by white circles for elliptical and lenticular (E) galaxies and white triangles for early-type spiral (S) galaxies.}
\label{diag}
\end{figure*}

In Fig.\ \ref{diag}, we plot the two emission line ratios for all galaxies with $m_{r}<$16 and $\geq 5 \sigma$ detections (i.e., the emission line fluxes are at least five times the uncertainty given by \citet{tre04}) of the $[$NII$] \lambda$6583, H$\beta$, and H$\alpha$ emission lines.  In the upper panel of the figure, we only include those galaxies with $\geq 3 \sigma$ detections of the typically weaker $[$OIII$] \lambda$5007 line.  For galaxies that have no significant $[$OIII$]$ emission but have $\geq 5 \sigma$ detections of the other three emission lines, we plot upper limits for $[$OIII$]$/H$\beta$  in the lower panel of Fig.\ \ref{diag}.  We also plot the theoretical upper limit for star-burst galaxies from \citet{kew01} as a dashed curve in both panels of Fig.\ \ref{diag}.  From these plots, it is apparent that the star-forming and AGN galaxies form two relatively distinct sequences and that even though they are consistent with star-burst galaxies, many of the galaxies that lie below the \citet{kew01} boundary may contain significant AGN components.  Since the vast majority of elliptical and lenticular galaxies have emission line ratios consistent with extended LINER-like emission \citep{phi86,gou94,sar06}, we do not want to use such a generous upper limit for the selection of our sample since our ultimate goal is to identify elliptical and lenticular galaxies that are most likely forming stars.  We have therefore adjusted the curve from \citet{kew01} to provided a more conservative upper limit which is given by

\begin{equation}
\mbox{log } \left ( \frac{[\mbox{OIII}] \lambda 5007}{\mbox{H} \beta} \right ) = \frac{0.61}{\mbox{log }([\mbox{NII}] / \mbox{H} \alpha) - 0.12} + 1.14
\label{diageq}
\end{equation}
This upper limit is plotted as a solid curve in both panels of Fig.\ \ref{diag} and is similar to other empirically derived limits such as that used by \citet[][see Fig.\ \ref{diag}]{kau03a}.\par

\begin{figure*}
\includegraphics[scale=0.9]{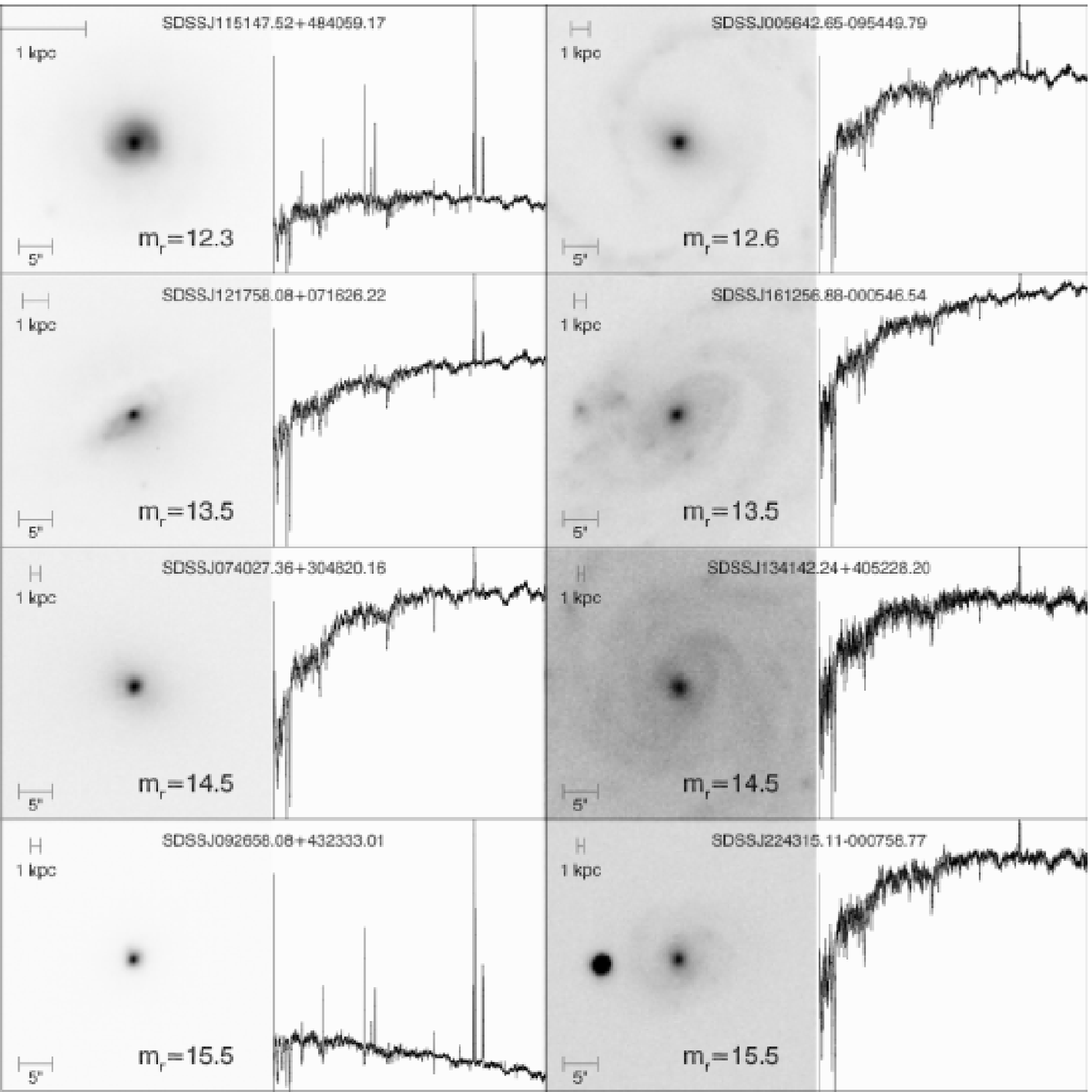}
\caption{Spectra (log F$_{\lambda}$ versus log $\lambda$) from 3800 to 8000 $\mbox{\AA}$ and $g$-band images of four SFE E galaxies (left, see \S 2) and four SFE S galaxies (right) chosen at four different $r$-band magnitudes ($m_{r} \approx$12.5, 13.5, 14.5, and 15.5).}
\label{exsfe}
\end{figure*}

\subsection{Identifying early-type galaxies}
With the boundary given by equation (\ref{diageq}), we can select a sample of star-forming galaxies within which we can search for elliptical and lenticular galaxies.  Using this boundary, we find $\sim$11,000 star-forming galaxies with $m_{r}<$16.  Morphologically classifying such a large sample by eye is impractical.  Therefore, we must isolate those star-forming galaxies that are most likely to be early-type galaxies so that we may obtain a more manageable number of galaxies for visual inspection to eliminate spiral contamination.  We have chosen to do this using a combination of the Sersic indexes, $n_{s}$, and the axis ratios, $b/a$, for the galaxies.  The Sersic indexes are taken from the Sersic fits performed by \citet{bla03b} to surface brightness profiles measured from the $i$-band images by the SDSS photometric pipeline using circular apertures.  Tests performed by \cite{bla03b} indicate that these fits perform reasonably well down to the magnitude limit of the Main sample for galaxies with $b/a ^{>}_{\sim}$0.5.  For the quantity $b/a$, we used values from the best fitting de Vaucouleurs profile as determined by the SDSS photometric pipeline using the $i$-band images which take the size of the seeing disk into account \citep[see ][]{sto02}.  The fits are performed assuming that the isophotal axis ratio does not change with radius, and consequently, one value of $b/a$ is solved for for each galaxy.  To determine the best combination of these two parameters, we have used the catalogue of 1875 morphologically classified SDSS galaxies compiled by \citet{nak03}.  We adjusted the imposed limits for $n_{s}$ and $b/a$ so that the largest number of the elliptical galaxies in the \citet{nak03} catalogue would be retained while at the same time, the largest number of galaxies of type Sa or later would be rejected.  The criteria we derived were $n_{s}>$3.5 and $b/a>$0.7.  From the \citet{nak03} catalogue, 67\% of the galaxies that pass these two cuts are E or E/S0 galaxies, 16\% are S0 or S0/a galaxies, 16\% are of type Sa or later, and 1\% are irregular or peculiar galaxies.  About 90\% of all the elliptical galaxies within the \citet{nak03} catalogue meet the two criteria.  We note that 25\% of the galaxies within the \citet{nak03} catalogue that have $n_{s}>$3.5 are spiral galaxies, implying that the additional cut based on axis ratio is useful.  We also note that since the Sersic fits are more reliable for galaxies with larger axis ratios and that the $b/a$ values taken from the SDSS pipeline are more appropriate for galaxies with $n_{s} \sim$4, the two criteria we have chosen imply that the measured values for $n_{s}$ and $b/a$ are reliable.  Hereafter, when we refer to galaxies as "early-type", we are referring to those galaxies with $n_{s}>$3.5 and $b/a>$0.7; those star-forming galaxies that meet these criteria are referred to as star-forming early-type, or "SFE" galaxies.\par 
Among the star-forming galaxies from DR4 that have $m_{r}<$16, 752 have $n_{s}>$3.5 and $b/a>$0.7, a reasonable number of SFE galaxies for which visual inspection of the images can be done in a practical amount of time.   We have examined the $g$-band images of these galaxies and have put them into three categories, (i) elliptical and lenticular galaxies, (ii) galaxies that have prominent disk components, spiral arms, or disks with "lumpy" structure indicative of H{\sc ii} regions (i.e., likely spiral galaxies), and (iii) peculiar, irregular, or merging galaxies as well as H{\sc ii} regions within nearby galaxies incorrectly identified as individual galaxies by the SDSS pipeline.  For convenience, we hereafter refer to these three groups as E, S, and P.  The numbers of SFE galaxies within these groups are summarised in Table \ref{galnum}.  The $g$-band images and spectra of four SFE E galaxies and four SFE S galaxies chosen at four different $r$-band magnitudes ($m_{r} \approx$12.5, 13.5, 14.5, and 15.5) are displayed in Fig.\ \ref{exsfe}.  It should be noted that we have not performed a detailed morphological classification of our SFE galaxies.  For instance, all galaxies with dominant spheroidal components and no prominent disks or spiral structures were placed within the E category, regardless of any slightly atypical features such as off-centre nuclei (e.g., see the g-band image of SDSS J121758.08+071626.22 in Fig.\ \ref{exsfe}), and galaxies within the S category were not broken up into subclasses such as Sa, Sb, Sc, etc.\par
\begin{figure}
\includegraphics[scale=0.42]{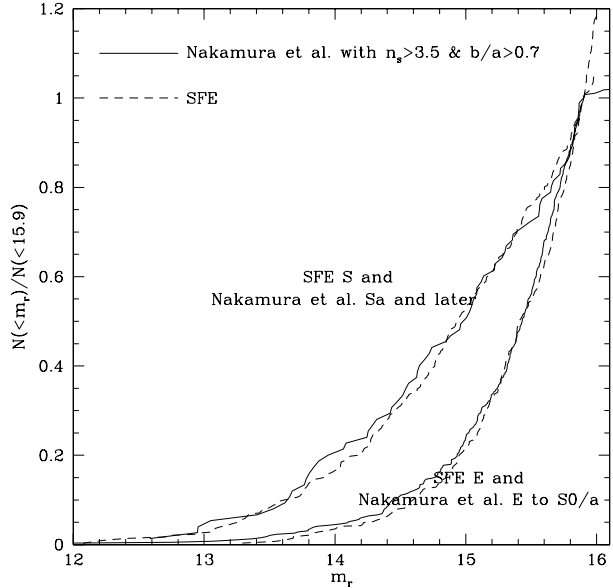}
\caption{A comparison of the visual classification of our SFE galaxies to that of the \citet{nak03} sample of SDSS galaxies.  The cumulative distributions of the $r$-band magnitude for E, E/S0, S0, and S0/a galaxies from the \citet{nak03} catalogue that have $n_{s}>$3.5 and $b/a>$0.7 is represented by the lower solid line.  The distribution for the Sa or later-type galaxies from the \citet{nak03} catalogue that have $n_{s}>$3.5 and $b/a>$0.7 is plotted as the upper solid line.  Similar cumulative distributions are also plotted for SFE E (see \S 2) galaxies (lower dashed line), and for SFE S galaxies (upper dashed line).  Each curve has been normalised to the number of galaxies at $m_{r}$=15.9, which is similar to the $r^{\prime}$-band (i.e., in the older SDSS photometric system) magnitude limit of the \citet{nak03} catalogue.  The fact that the shapes of the curves for our SFE E and SFE S samples match those of the curves for the corresponding \citet{nak03} sub-samples implies that our visual inspection has performed as well as that of \citet{nak03} as a function of apparent magnitude.}
\label{mrdist}
\end{figure}

While Fig.\ \ref{exsfe} demonstrates that our visual inspection has worked well for those cases, it does little to reassure us that our morphological classification is consistent over the entire range in redshifts/apparent size inhabited by our sample galaxies.  We can roughly test the quality/reliability of our visual classifications by comparing our visual inspection as a function of apparent magnitude to that of \citet{nak03} to at least see if our classifications are as consistent as theirs are as a function of distance.  To do this, we have computed the cumulative distributions of $m_{r}$ for all elliptical and lenticular galaxies (type S0/a or earlier) from the \citet{nak03} sample that have $n_{s}>$3.5 and $b/a>$0.7 and for all spiral galaxies (type Sa or later) that also have $n_{s}>$3.5 and $b/a>$0.7.  These distributions are plotted in Fig.\ \ref{mrdist} along with the same distributions for our SFE E and SFE S galaxies.  Each distribution has been normalised by the number of galaxies with $m_{r}<$15.9 since the \citet{nak03} sample is significantly incomplete for magnitudes fainter than this limit.  From Fig.\ \ref{mrdist}, it is apparent that the shapes of the distributions for our SFE E and SFE S galaxies match those of the corresponding \citet{nak03} sub-samples.  This implies that while we have no direct test of how well our morphological classification works with redshift, we can at least be sure that our visual inspection has performed as well as that of \citet{nak03} as a function of apparent magnitude.\par
Since we are reasonably confident in our morphological classifications, we have used these classifications to eliminate spiral galaxies (i.e., SFE S galaxies) from our analysis.  As discussed in \S 1, it is more likely that elliptical and lenticular K+A galaxies result from episodes of star formation within morphologically similar galaxies while K+A spiral galaxies may result from either a similar scenario or from a truncation of star formation by processes such as gas stripping.  This implies that while the SFE E galaxies are excellent candidates for the progenitors of morphologically similar K+A galaxies, it is substantially more difficult to tie the SFE S galaxies to spiral K+A galaxies in a similar way.  This is made even more difficult by the fact that the 3 arcsec SDSS spectrograph aperture may miss significant amounts of disk star formation in nearby galaxies (see Fig.\ \ref{exsfe}), implying that many spiral galaxies may be misclassified as K+A using their SDSS spectra.  For these reasons, the remainder of this paper will focus on the SFE E galaxies and how they relate to morphologically similar K+A galaxies.\par

\begin{table}
\centering
\caption{Galaxy Categories}
\begin{tabular}{lr}
\hline
Category & Number \\
\hline
All SDSS galaxies with $m_{r}<$16 & 39,621 \\
Early-type with $m_{r}<$16      & 10,870 \\
S/N$>$5 for H$\beta$, H$\alpha$, and $[$NII$]$ & 16,282 \\
S/N$>$5 for H$\beta$, H$\alpha$, $[$NII$]$, and S/N$>$3 for $[$OIII$]$ & 15,091 \\
AGN with $m_{r}<$16 & 1,754 \\
Star Forming with $m_{r}<$16 & 11,417 \\
SFE   				 &     752 \\
$\;\;\;\;\;$SFE E				 &     335 \\
$\;\;\;\;\;$SFE S				 &     402 \\
$\;\;\;\;\;$SFE P				 &       15 \\
K+A, $m_{r}<$16 		 &     435 \\
$\;\;\;\;\;$K+A, $m_{r}<$16, S/N$>$3 for H$\alpha$ & 220 \\
$\;\;\;\;\;$K+A E				 &     253 \\
$\;\;\;\;\;$K+A S				 &     142 \\
$\;\;\;\;\;$K+A P				 &       40 \\
\hline
\end{tabular}
\label{galnum}
\end{table}

\begin{figure}
\includegraphics[scale=0.42]{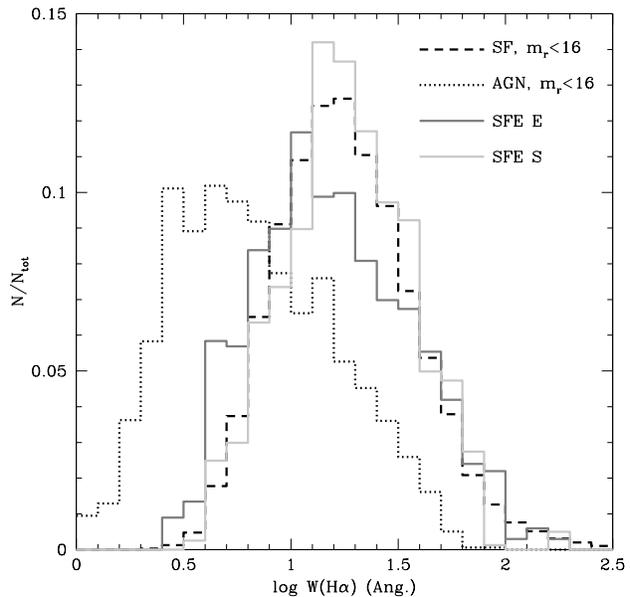}
\caption{The distributions for the H$\alpha$ emission line equivalent width, $W(H\alpha)$, for all star forming galaxies with $m_{r}<$16 (dashed line) and for SFE E galaxies (solid line).  For comparison, the $W(H\alpha)$ distribution for AGN galaxies with $m_{r}<$16 (dotted line) is also plotted where we consider all galaxies with values of $[$OIII$]$/H$\beta$ greater than the upper limit for star-burst galaxies given by \citet{kew01} illustrated in Fig.\ \ref{diag}.  For each histogram, $N_{tot}$ refers to the total number of galaxies within the corresponding sub-sample.}
\label{wha}
\end{figure}

To roughly compare the range of star formation rates (SFRs) occupied by our SFE E galaxies with that for "normal" star forming galaxies, we have plotted the H$\alpha$ emission line equivalent width distributions for these galaxies in Fig.\ \ref{wha}.  Since the H$\alpha$ luminosity is a direct measure of the SFR \citep{ken98}, the H$\alpha$ equivalent width, $W(H\alpha)$, is a good tracer of the SFR per unit stellar mass.  In Fig.\ \ref{wha}, we have plotted separately the $W(H\alpha)$ distributions for all star forming galaxies with $m_{r}<16$ and for SFE E galaxies.  From these distributions, we find that in terms of $W(H\alpha)$, our SFE E galaxies are relatively similar to more typical star forming galaxies with slightly (a factor of $\sim$1.1 times) lower values of $W(H\alpha)$ on average.  For comparison, we have also plotted the $W(H\alpha)$ distribution for AGN galaxies, which we take to be all galaxies with $m_{r}<$16 that have values of $[$OIII$]$/H$\beta$ that are greater than the \citet{kew01} upper limit for star-burst galaxies (see Table \ref{galnum}) displayed in Fig.\ \ref{diag}.  The typical value of $W(H\alpha)$ for these galaxies is factor of $\sim$3 lower than what was found for star forming galaxies, significantly lower than what was found for SFE E galaxies.  This implies that in terms of their levels of H$\alpha$ emission, SFE E galaxies are more consistent with typical star forming galaxies than with AGN.  It should also be noted that the H$\alpha$ equivalent widths for the SFE E galaxies are substantially higher than what can be produced by photoionisation by post-asymptotic giant branch stars (PAGB) within post-star-burst or older, quiescent stellar populations which ranges from about 0.3 to 3 $\mbox{\AA}$ according to the models of \citet{bin94}.  This is in contrast with typical early-type galaxies where the level of H$\alpha$ emission is usually consistent with both LINER galaxies and model predictions of photoionisation by PAGB stars \citep[e.g.,][]{gou94,bin94,mac96}.

\subsection{Identifying K+A galaxies}
Since our major goal is to explore the possibility that our SFE E galaxies will become morphologically similar K+A galaxies, we need to have a sample of these objects as well.  Among all Main galaxies with $m_{r}<16$, we have identified K+A galaxies of all morphological types and have used visual inspection of the $g$-band images to identify E and S0 K+A galaxies.  The first step in this process was to choose a definition for K+A galaxies.  We have adopted a K+A definition that is very similar to that used by \citet{qui04} to identify K+A galaxies using SDSS spectra.  The main difference is that we have chosen to use the strength of the H$\delta$ absorption line instead of the quantity $A/K$ used by \citet{qui04}.  The quantity $A/K$ is essentially the ratio of the flux between 3800 and 5400 $\mbox{\AA}$ due to the recently (within $\sim$1 Gyr) formed stellar component (i.e., the component that resembles a main sequence A star) to that due to the more quiescent component (i.e., the component that resembles a K giant star).  We have opted to use the strength of the H$\delta$ absorption line instead of $A/K$ because the $H\delta$ absorption equivalent width is relatively insensitive to dust extinction when compared to the ratio $A/K$ which is based on the shape of a galaxy's optical spectrum over a range of 1600 $\mbox{\AA}$ in wavelength.\par
\begin{figure*}
\includegraphics[scale=0.75]{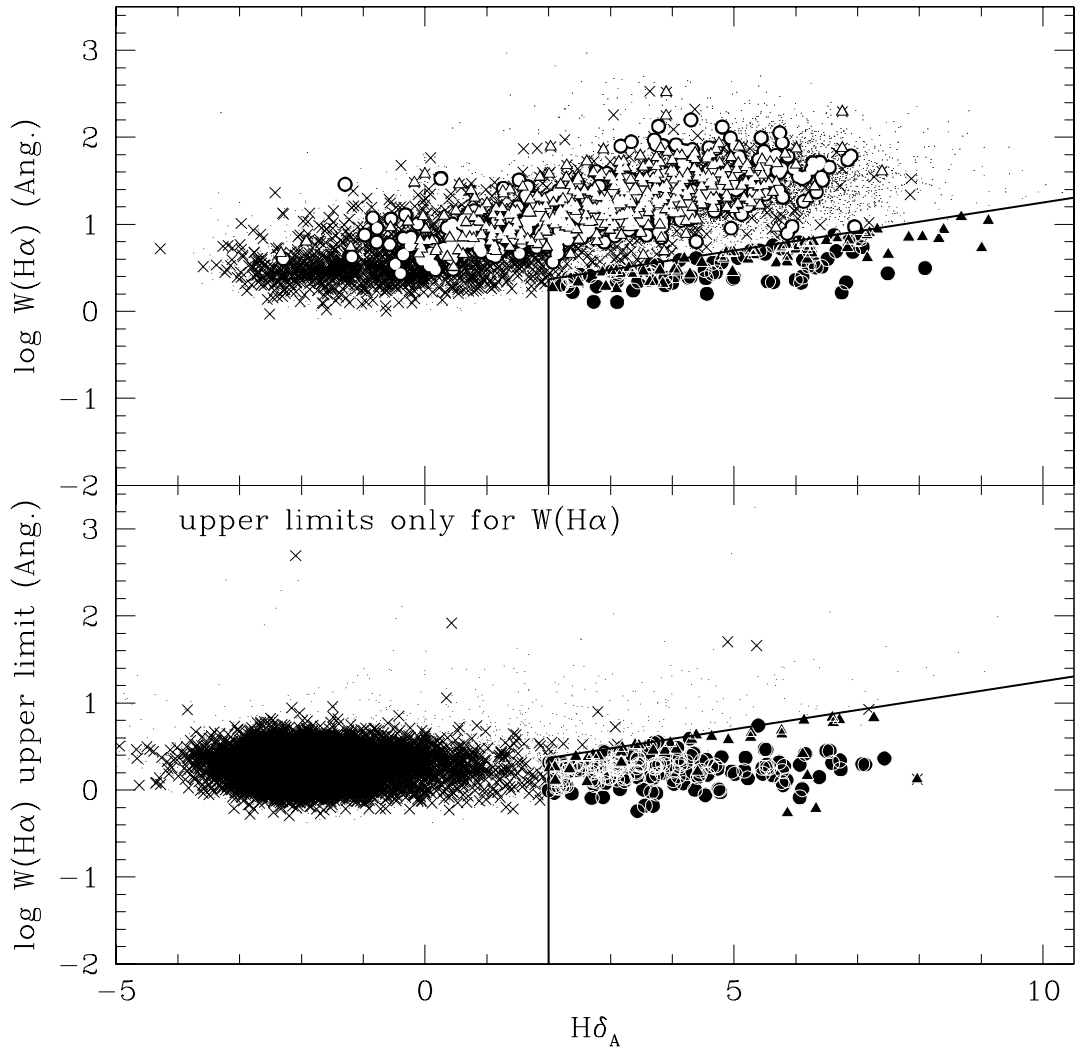}
\caption{The H$\alpha$ emission line equivalent width, $W(H\alpha)$ ({\it emission} is positive; see \S 2.3), versus the strength of the H$\delta$ absorption line, $H\delta_{A}$ ({\it absorption} is positive; see \S 2.3), for galaxies with $\geq 3 \sigma$ detections of H$\alpha$ and $m_{r}<$16 (upper).  For galaxies with no significant detection of H$\alpha$, the upper limit for $W(H\alpha)$ is plotted in the lower panel.  The point styles and are the same as those used in Fig.\ \ref{diag} with elliptical and lenticular K+A galaxies represented by black circles and spiral K+A galaxies represented by black triangles.  In both panels, the adopted definition of K+A galaxies is illustrated by the solid lines (see \S 2.3).}
\label{classdef}
\end{figure*}

To develop our own similar K+A definition, we have chosen to use the spectral index H$\delta_{A}$ \citep{wor97} and the H$\alpha$ emission line equivalent width, $W(H\alpha)$.  The index $H\delta_{A}$ is designed to provide a measure of the equivalent width of the stellar H$\delta$ absorption line and is defined such that positive values imply {\it absorption}.  For the nebular H$\alpha$ emission line equivalent width, $W(H\alpha)$, we have chosen to associate positive values with {\it emission} because (i) such equivalent widths typically occupy a large range in values and it is more convenient to work with values of log $W(H\alpha)$ and (ii) we wish to be able to compare our efforts with similar work performed by previous authors who have used the same convention \citep[e.g.,][]{qui04}.  For both $H\delta_{A}$ and $W(H\alpha)$, we used the values measured by \citet{tre04} where the values of H$\delta_{A}$ were corrected by \citet{tre04} for emission lines that originate from the individual galaxies and in some cases, from foreground sky emission.  In the upper panel of Fig.\ \ref{classdef}, we plot $W(H\alpha)$ versus H$\delta_{A}$ for galaxies with $m_{r}<$16 and $\geq 3 \sigma$ detections of the H$\alpha$ emission line.  For galaxies with no significant H$\alpha$ detection, we plot upper limits for $W(H\alpha)$ versus H$\delta_{A}$ in the lower panel of Fig.\ \ref{classdef}.  In developing a definition for K+A galaxies, we sought to exclude actively star-forming and quiescent galaxies while including as many K+A galaxies as possible.  To exclude star-forming galaxies, we have empirically derived an upper limit for $W(H\alpha)$ as a function of H$\delta_{A}$ using the lowest observed values for $W(H\alpha)$ for star-forming galaxies as they are defined in Fig.\ \ref{diag}.  This is very similar to what was done by \citet{qui04} who used an upper limit for $W(H\alpha)$ which varied linearly with $A/K$.  Our adopted upper limit is plotted as a solid line in both panels of Fig.\ \ref{classdef}.  Formally, our K+A definition requires log $W(H\alpha)<0.11H\delta_{A}+0.15$ and H$\delta_{A}>2 \mbox{ \AA}$; this second requirement is designed to exclude quiescent galaxies.\par
Using the K+A definition illustrated in Fig.\ \ref{classdef}, we found 435 K+A galaxies within DR4 with $m_{r}<$16 which make up $\sim$1\% of all DR4 galaxies with $m_{r}<$16.  This is very similar to the K+A galaxies within the \citet{qui04} sample which make up roughly 1\% of galaxies from the SDSS Main sample.  We note that the relative number of galaxies in either our K+A sample or the \citet{qui04} sample is approximately an order of magnitude higher than has been found using more strict definitions of K+A galaxies \citep[e.g.,][]{zab96,got03}.  We also note that roughly half of our K+A galaxies have $>3\sigma$ detections of H$\alpha$ (see Table \ref{galnum} and Fig.\ \ref{classdef}) while only about 6\% have $>3\sigma$ detections of H$\delta$ emission.  We further note that both our K+A selection criteria and those of \citet{qui04} do not include any requirement based on the flux of the $[$OII$]\lambda$3727 doublet, unlike some previous authors \citep[e.g.,][]{zab96,got03}.  Several authors have noted a significant fraction of galaxies with strong Balmer absorption and weak or no H$\alpha$ emission do have significant levels of $[$OII$]$ emission \citep[e.g.,][]{liu95,zab96,got03,qui04}.  About 18\% of our K+A galaxies have $\geq 3 \sigma$ detections of $[$OII$]$ emission with a typical equivalent width of $\sim10 \mbox{ \AA}$.  Among those K+A galaxies with detected H$\alpha$ and $[$OII$]$ emission, the ratio of $[$OII$]$/H$\alpha$ is consistent with what has been found for more typical early-type galaxies \citep{gou94,bin94}.  This implies that the $[$OII$]$ emission is not indicative of residual star formation and is more consistent with LINER-like emission \citep[e.g.,][]{gou94,yan06}, possibly associated with photoionisation by PAGB stars \citep{bin94,gou99}.\par
To identify elliptical and lenticular K+A galaxies we have used visual inspection of $g$-band images.  Since there are only 435 K+A galaxies with $m_{r}<16$, as opposed to 11,417 star forming galaxies, it was not necessary to apply the same morphological selection criteria to provide a smaller, more manageable list of images to inspect. We have therefore not required that the K+A galaxies have $n_{s}>3.5$ and $b/a>0.7$.  After inspecting the $g$-band images of the K+A galaxies, we put them into the same three morphological categories as were used for the SFE galaxies, E, S, and P.  The numbers of K+A galaxies within each of these groups are summarised in Table \ref{galnum}.  These morphological classifications provided us with a sample of 253 elliptical and lenticular K+A galaxies with which we can compare our SFE E galaxies.\par

\section{Early-type star forming and early-type K+A galaxies}
\begin{figure}
\includegraphics[scale=0.42]{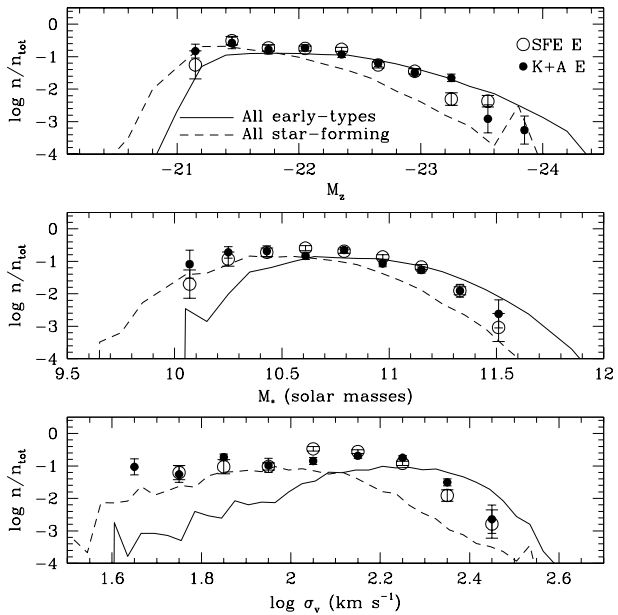}
\caption{Galaxy number density as a function of M$_{^{0.1}z}$ (the absolute $z$-band magnitude K-corrected to z$=$0.1, the median redshift of the Main sample; upper), stellar mass, M$_{\ast}$ (middle), and velocity dispersion, $\sigma_{v}$ (lower), for SFE E (open circles) and K+A E (filled circles).  We also plot the same for all star-forming galaxies (dashed line; see \S 2.1) and all early-type galaxies (solid line; see \S 2.2) with $m_{r}<$16 so that the shapes of the SFE and K+A distributions may be compared to the shapes of these distributions.  In all panels, the number density in each bin is divided by the total number density since our relatively complex selection criteria preclude any meaningful comparison of absolute number densities among the sub-samples.}
\label{functions}
\end{figure}

If our SFE E galaxies will in fact become K+A E galaxies, then they must have a few basic properties in common with the galaxies within the K+A E sample.  To check this, we first used the 1/$V_{max}$ method to estimate the number density of SFE E and K+A E galaxies with z$>$0.05 and $m_{r}>$13 within different bins of absolute $^{0.1}z$-band (i.e., K-corrected to a redshift of 0.1) magnitude, M$_{^{0.1}z}$, stellar mass, M$_{\ast}$, and stellar line-of-sight (LOS) velocity dispersion, $\sigma_{v}$; the results are plotted in Fig.\ \ref{functions}.  For these calculations, and for all other distance dependent calculations throughout this paper, we assumed ($\Omega_{m}, \Omega_{\Lambda}, h$)=(0.3,0.7,0.7).  The lower redshift limit of 0.05 was used to minimise the effects of aperture bias resulting from the use of a fixed angular aperture \citep[i.e., the SDSS spectrograph 3 arcsec fibre aperture; ][]{gom03}.  The lower magnitude limit of 13 was implemented to exclude the most nearby galaxies.  This was done because nearby galaxies with bright cores will not be targeted for spectroscopy by the SDSS pipeline to prevent crosstalk among fibres \citep{str02}, implying that the 1/$V_{max}$ method will provide an insufficient completeness correction for these objects.  The values for M$_{^{0.1}z}$ and M$_{\ast}$ were taken from \citep{kau03} and include their K-corrections and extinction corrections.   To obtain measurements of $\sigma_{v}$, we have used the IDL routine {\it vdispfit} written by D. Schlegel.  This routine determines the best-fitting velocity dispersion and the 1$\sigma$ error in that dispersion by cross-correlating each spectrum with several template spectra that have been broadened by various Gaussian velocity distributions while masking regions of the spectrum that may contain emission lines.  For each of the two groups, the distributions have been normalised by the total estimated number density.  This was done because the relatively complex selection process used to cull each of the sub-samples precludes any meaningful comparison of the actual number densities between the two groups, and we are therefore only interested in comparing the shapes of the sub-samples' distributions.\par
We have also computed the number density of all star-forming (see \S 2.1) and all early-type (see \S 2.2) galaxies from our parent sample with z$>$0.05 and 13$<m_{r}<$16 within bins of M$_{^{0.1}z}$, M$_{\ast}$, and $\sigma_{v}$.  So that the shapes of these distributions can be compared with those of the SFE and K+A galaxies, we have plotted the results in Fig.\ \ref{functions} with each distribution normalised by its total estimated number density.  From Fig.\ \ref{functions}, it is apparent that for all three quantities, the distribution functions for SFE E and K+A E galaxies both lie somewhere in between those for normal star-forming galaxies and those for typical early-type galaxies.  This implies that each of these types of galaxies is on average more massive than the typical star-forming galaxy, but less massive than the average early-type galaxy.  The shapes of the distributions for all three properties appear essentially identical for both SFE E and K+A E galaxies.  This implies that the galaxies from both groups have about the same stellar and dynamical masses on average.\par

\begin{figure}
\includegraphics[scale=0.42]{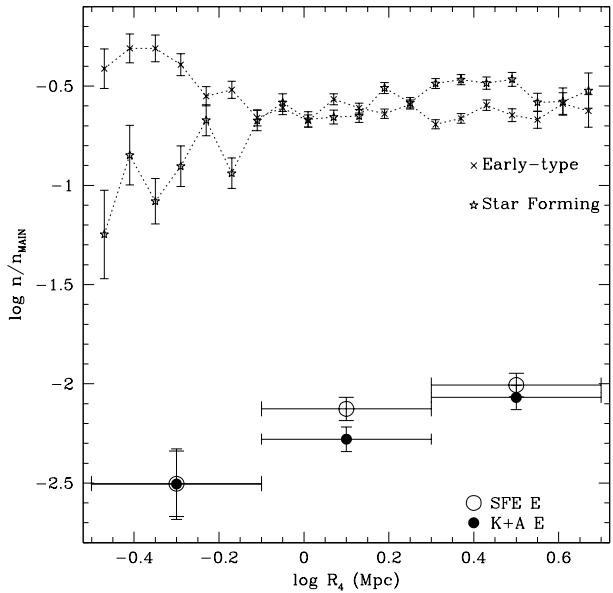}
\caption{Within bins of the distance to the fourth nearest Main galaxy within 1200 km s$^{-1}$ with $M_{r}<$-20, $R_{4}$, the fraction of the total number density of Main galaxies attributed to star-forming (stars), all early-type ($\times$'s), SFE E (open circles), and K+A E (filled circles) galaxies.  All galaxies used have 13$<m_{r}<$16 and 0.05$<$z$<$0.079 (see \S 3).}
\label{env}
\end{figure}

The results above imply that our SFE E galaxies have roughly the same masses as K+A E galaxies, but are they located in the same environments?  To address this question, we have quantified the environment around each galaxy by measuring the projected distance, $R_{4}$, to the fourth nearest Main galaxy with a radial velocity separation $<$1200 km s$^{-1}$ and with an absolute $r$-band magnitude brighter than -20.  Given the Main sample limiting magnitude of 17.77, this allows us to measure $R_{4}$ for all galaxies out to a redshift of 0.079.  We have again used the 1/$V_{max}$ method to compute the number density of all star-forming, early-type, SFE E, and K+A E galaxies with 13$<m_{r}<$16 and 0.05$<$z$<$0.079, this time within different $R_{4}$ bins.  In Fig.\ \ref{env}, we plot the fraction of the total number density of Main galaxies with 13$<m_{r}<$16 and 0.05$<$z$<$0.079 contained within each galaxy category for each of these bins.  As may be expected, the fraction of star-forming galaxies drops substantially at lower values of $R_{4}$ whereas the fraction of early-type galaxies does the exact opposite.  The fraction of both SFE E and K+A E galaxies is higher for the largest $R_{4}$ bin, implying that in terms of environment, these galaxies have more in common with typical star-forming galaxies than most early-type galaxies.  This is consistent with what was found for the K+A galaxy samples of \citet{hog05} and \citet{got05} as well as the somewhat more evolved "E+F" galaxies from the sample of \citet{nol07}.  It should also be noted that the shapes of the trends between the fractions of SFE E and K+A E galaxies and $R_{4}$ are quite similar, implying that the two types of galaxies typically reside within similar environments.  Combined with the findings of the authors cited above, this indicates that it is plausible that our SFE E galaxies lie along a common evolutionary sequence with both K+A and E+F galaxies.

\section{Modelling the stellar continua of star-forming elliptical and lenticular galaxies}
\subsection{Basic assumptions}
The fact that our SFE E galaxies make up a relatively small fraction ($\sim$3\%) of all early-type galaxies implies that the current star formation within these galaxies is likely short-lived.  Given their morphological similarity to the majority of K+A galaxies which are roughly just as rare among all early-type galaxies, it seems likely that SFE E galaxies will evolve into early-type K+A galaxies. The fact that the mass distributions and environments of our samples of SFE E and K+A E galaxies are similar is consistent with this conclusion.  In this section we will explore the likely evolutionary link between our SFE E galaxies and early-type K+A galaxies in more detail.  We will discuss our estimates of the magnitude and time-scale of the star formation in the early-type galaxies, and the subsequent spectral evolution of the SFE E galaxies. We will do this from model fits to their stellar spectra which utilise the population synthesis models of \citet{bru03}.  We will also discuss how these fits were used to determine under which conditions the galaxies will appear as K+A galaxies with spectroscopic properties as defined by us earlier (see \S 2.3 and Fig.\ \ref{classdef}).\par
To construct models that can be fit to the SFE E galaxy spectra, the main assumption that we have made is that the SFE E galaxies consist of two basic stellar components, an older pre-existing population and a new population that is currently being formed.  In fact, \citet{bal05} concluded that the stellar H$\delta$ absorption equivalent widths and near-IR colours of many K+A galaxies are best explained by star formation histories where relatively new (within the last $\sim$1 Gyr) stars were formed among existing older stellar populations.  We note that no explicit assumption was made regarding the formation mechanism (e.g., monolithic collapse or major mergers) or formation epoch for the assumed underlying older stellar population; we only assumed that the gross spectral properties of this population can be well approximated with an instantaneous burst model at a relatively large age.  We have also made the simplifying assumption that the star formation within the SFE E galaxies exponentially decreases with time.  Since the stellar continua alone cannot be used to effectively constrain the e-folding time, $\tau$, for this scenario, we have chosen to perform four separate fits for each spectrum, assuming $\tau = 100$ Myr, 300 Myr, 1 Gyr, and $\infty$ (i.e., a constant SFR).  To further simplify our modelling efforts, we have chosen to only use \citet{bru03} models that assume the initial mass function of \citet{cha03}.  These assumptions allow us to use the \citet{bru03} instantaneous burst models to compute suites of model spectra for the older population and the exponentially decreasing and constant SFR models to compute the same for the younger stellar components.  To this end, we have constructed a suite of 1100 spectra for each of these models using five different metallicities (Z$=$0.0004, 0.004, 0.008, 0.02, and 0.05) and 220 times steps ranging from $10^{5.1}$ to $10^{10.3}$ yr.  Each of the model spectra was normalised by dividing by the total stellar mass at the appropriate time step which includes the \citet{bru03} computation for the amount of stellar mass loss.
\subsection{Parameter constraints}
Our basic approach for estimating the age of the younger population within each SFE E galaxy was to find the best fitting linear combination of an instantaneous past burst spectrum and one of the exponentially decreasing SFR spectra from the five suites of model spectra.  The coefficients in this combination are essentially the stellar masses of the two component populations.  However, in practice, we must recognise that (i) the stellar mass, age, and metallicity are not the only factors that affect the shapes of the spectra and (ii) the effects of different properties on the spectral shapes may be degenerate (e.g., age and metallicitiy).  The two additional properties that must be considered when finding the best fitting model spectrum are the amount of internal dust extinction and the velocity dispersion.  To simplify matters, we assumed one velocity dispersion for the combined model spectrum, the dispersion obtained for the actual galaxy spectrum using {\it vdispfit} (see \S 3).  We have also made the simplifying assumption that the old and new populations suffer the same amount of dust extinction.  This assumption is relatively good if the star forming regions occupy an area greater than or equal to the area covered by the SDSS spectrograph fibre aperture.  Narrow-band imaging of a few or our SFE E galaxies suggests that this is indeed the case \citep{hel05}.  For each spectrum, we therefore must fit for several different parameters and we must constrain each of these as strictly as possible to overcome the degeneracy between the age of the younger population (the main parameter we are interested in) and any other parameter.  We have done this in the following way:
\begin{enumerate}
\item {\it Metallicity}:  We have good estimates for the oxygen abundance for the ionised gas within each SFE E galaxy from the work of \citet{tre04}.  It is likely that the ionised gas is associated with the newly forming stars and that their metallicities are similar.    However, over time, the metallicity of the newly formed stellar population may lag behind that of the ionised gas as the gas is continually enriched.  This implies that for younger populations with relatively large ages, the O/H abundance from \citet{tre04} may overestimate the true abundance of the stars.  Even so, placing strict limits on the stellar metallicity is crucial to providing reasonable age estimates for the younger stellar populations and the \citet{tre04} values are the only independent metallicity estimates available. We have therefore chosen to constrain the metallicity for the younger stellar population to be within the 68\% confidence interval of the O/H value given by \citet{tre04}, and caution the reader that for galaxies with relatively large age estimates for their younger populations, the O/H abundances may be somewhat inflated.\par
For the older population, we do not have a comparable constraint for the stellar metallicity.  While accurately estimating the age of the older population is not crucial to our main goal, we may partially constrain the older population's metallicity using the spectral index [MgFe]$^{\prime}$ \citep{tho03}.  While not independent of age, [MgFe]$^{\prime}$ is more strongly affected by metallicity and has the added advantage of being nearly independent of the abundance of $\alpha$-elements.  The latter is relevant since many early-type galaxies are known to be $\alpha$-enhanced relative to the sun \citep[e.g., ][]{tho03} and the \citet{bru03} models assume solar values for the abundances of heavier elements relative to one another.  According to the \citet{bru03} models, an instantaneous burst population that is older than $\sim$1 Gyr will have a larger value of [MgFe]$^{\prime}$ than an actively forming population, regardless of the new population's age.  This implies that the measured value of [MgFe]$^{\prime}$ for the actual galaxy spectrum provides a lower limit for [MgFe]$^{\prime}$ for the older population.  The \citet{bru03} models also imply that [MgFe]$^{\prime}$ will not exceed 5 $\mbox{\AA}$.  We therefore constrain the value of [MgFe]$^{\prime}$ for the older population to be $<5\mbox{ \AA}$ and to be greater than [MgFe]$^{\prime}_{obs}-1\sigma$ where [MgFe]$^{\prime}_{obs}$ and $\sigma$ are the measured values for [MgFe]$^{\prime}$ and their uncertainties taken from the SDSS spectroscopic pipeline.
\item {\it Dust extinction}:  Using the observed ratio of H$\alpha$ to H$\beta$, we can obtain a good estimate of the internal dust extinction suffered by the ionised gas which can be used to constrain the amount of internal dust extinction.  To estimate the extinction using the Balmer decrement, we must assume an intrinsic value for H$\alpha$/H$\beta$ which is dependent upon temperature.  To determine the best temperature to use, we have estimated the electron temperature using the [OII], [OIII], and H$\beta$ emission lines according to \citet{pil01} for each SFE E galaxy with $>3\sigma$ detections of all of these emission  lines.  The median temperature for these galaxies is 5500 K, for which the ratio of H$\alpha$ to H$\beta$ is 3.02 for case B recombination according to \citet{ost06}.  We used this assumed intrinsic ratio to compute the colour excess for the ionised gas, $E(B-V)_{gas}$, for each SFE E galaxy using the extinction curve of \citet{odo94} and $R_{V}=3.1$.  We have also used the errors in the H$\alpha$ and H$\beta$ fluxes from \citet{tre04} to compute the $1\sigma$ uncertainty in $E(B-V)_{gas}$ and constrained the fitted values for this quantity to be within $\pm 1\sigma$ of the observed value.  When computing the model spectra, the dust extinction was applied to the model stellar monochromatic fluxes using the extinction "law" of \citet{cal01}.  This extinction law uses the quantity $E(B-V)_{gas}$ and takes into account the wavelength dependence of the dust extinction as well as scattering and the geometrical distribution of the dust relative to the stars.
\item {\it Stellar mass and age}:  While we have no reliable estimates for the stellar masses of the older and younger stellar populations within each SFE E galaxy, we do have estimates for the total masses from \citet{kau03}.  However, it should be noted that \citet{kau03} measured stellar masses using linear combinations of \citet{bru03} model spectra, implying that they do not provide stellar mass measurements that are independent of our own fitting procedure.  We therefore used the \citet{kau03} value as an initial guess for the total stellar mass of each galaxy, but allowed the fitted value to vary by $\pm$1 dex from this value.  The fraction of the total stellar mass attributed to the younger population, $f_{y}$, was used to quantify the relative contribution of each population to the total stellar mass and was only constrained to be $\leq 1$ and $>0$.  The age of the younger population was constrained to be older than 1 Myr while the age of the older population was constrained to older than 1 Gyr; both populations were not allowed to be older than 13 Gyr.
\end{enumerate}

\subsection{Execution and reliability}
\begin{figure*}
\includegraphics[scale=0.9]{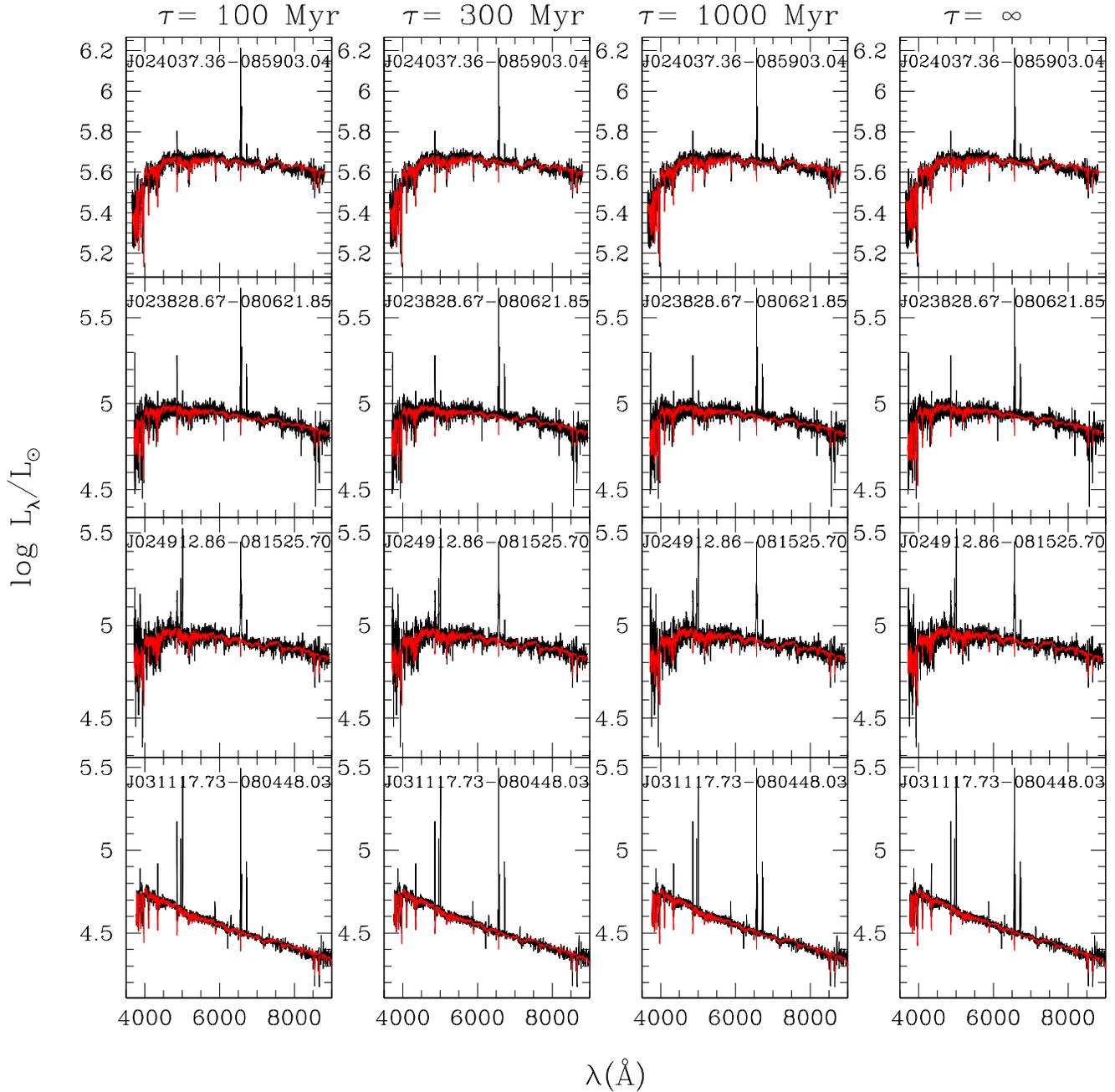}
\caption{Examples of the model spectra fits described in \S 4.2.  For each galaxy, we have plotted the SDSS spectrum (solid black) and the best fitting model spectrum (solid red) for SFR e-folding times of $\tau =$100 Myr, 300 Myr, 1 Gyr, and $\infty$ (i.e., constant SFR) in each column from left to right.  The monochromatic luminosity is plotted in units of $L_{\odot}=3.826\times 10^{33}$ ergs s$^{-1}$ and was computed assuming ($\Omega_{m}$,$\Omega_{\Lambda}$,$h$)$=$(0.3,0.7,0.7).}
\label{fitspec}
\end{figure*}

We have compiled a routine which will interpolate among the suite of instantaneous burst model spectra and one of the suites of exponentially decreasing model spectra described above to produce a composite spectrum for given values of eight parameters:  (1) the age of the older population, (2) the age of younger population, $t_{y}$, (3) the oxygen abundance of the younger population, O/H, (4) [MgFe]$^{\prime}$ for the older population, (5) the total stellar mass, $M_{\ast}$, (6) the fraction of the stellar mass contributed by the younger population, $f_{y}$, (7) the dust extinction/reddening for the ionised gas, $E(B-V)_{gas}$, and (8) the stellar velocity dispersion, $\sigma_{v}$.  Since we keep the velocity dispersion fixed at the value measured for the galaxy spectrum, this leaves seven free parameters.  We have used the IDL routine {\it mpfit} written by C. Marqwardt to determine the best fitting combination of these seven parameters for each SFE E galaxy that had an O/H measurement from \citet{tre04}, 278 galaxies in all.  This routine uses a standard Levenberg-Marquardt algorithm to minimise $\chi^{2}$, but allows for constraints to be placed on the free parameters and numerically computes the partial derivatives of the function used in the fitting with respect to the free parameters.  For each spectrum, regions of the spectrum with emission lines were not used in the fit and the spectrum was corrected for Galactic extinction using the $r$-band extinction from \citet{sch98}, the extinction curve of \citet{odo94}, and $R_{V}=3.1$.  For each fit, the free parameters were constrained as discussed above.  Examples of the SDSS spectra and the model fits for each value of $\tau$ can be seen in Fig.\ \ref{fitspec}.  Overall, the models appear to fit the data remarkably well (within about 0.01 dex on average), irrespective of what e-folding time was used.\par

\begin{figure}
\includegraphics[scale=0.42]{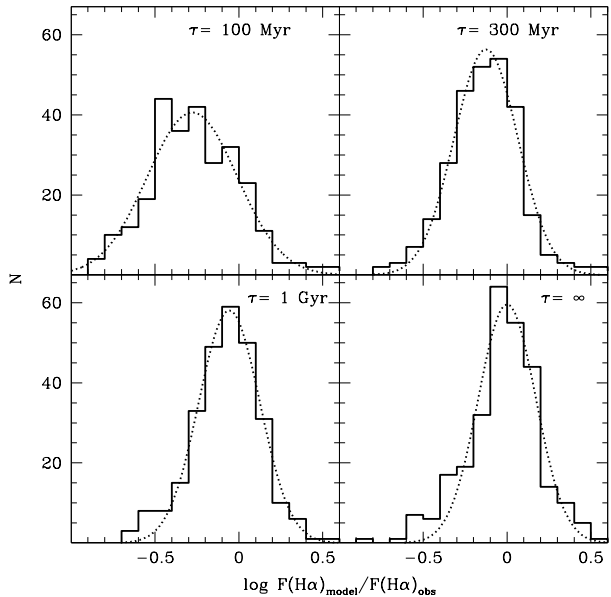}
\caption{For the 278 SFE E galaxies with gas-phase O/H values from \citet{tre04}, the distribution for the ratio of the  the H$\alpha$ flux predicted by each model fit assuming an electron temperature of 5500 K and case B recombination \citep{ost06} to the observed H$\alpha$ flux.  The distributions for each SFR e-folding time, $\tau$, are plotted separately with Gaussian fits to each plotted as dotted lines.  The means from these fits are -0.28 -0.13 -0.057 and 0.00031 dex and the standard deviations are 0.27, 0.19, 0.18, and 0.17 dex for $\tau =$100 Myr, 300 Myr, 1 Gyr, and $\infty$, respectively.}
\label{bcheck}
\end{figure}

Since the H$\alpha$ fluxes were not used to constrain the spectral fits, they provide a good opportunity to test how well these fits represent the SFE E galaxies.  Among other spectral properties, the \citet{bru03} models provide the the flux of hydrogen ionising photons, $Q(H^{0})$, at each time step for a given model, and we have used these values to determine $Q(H^{0})$ using the fitted ages, metallicities, and masses of the older and young populations for each galaxy and each value of $\tau$.  According to \citet{ost06}, for a temperature of 5500 K (see \S 4.2) and case B recombination, $L_{H\alpha} = 1.46\times 10^{-12} Q(H^{0})$ ergs s$^{-1}$.  We have used this equation with the fitted values of $E(B-V)_{gas}$ to compute the model predictions for the observed H$\alpha$ fluxes.  In Fig.\ \ref{bcheck}, we have displayed the distribution of the ratios of the model-predicted fluxes to the observed fluxes for each value of $\tau$.  Gaussian fits to these distributions estimate the means to be -0.28 -0.13 -0.057 and 0.00031 dex and the standard deviations to be 0.27, 0.19, 0.18, and 0.17 dex for $\tau =$100 Myr, 300 Myr, 1 Gyr, and $\infty$, respectively.  By far, the constant SFR ($\tau = \infty$) models provide the best statistical match to the observed H$\alpha$ fluxes with ratios of model-predicted to observed fluxes that range from roughly 0.7 to 1.5 with a median near unity.  We note that this result somewhat contradicts other similar studies which estimate that the typical e-folding time for galaxies similar to our SFE E galaxies is between 50 and 150 Myr \citep{kav07,sch07}.  However, we also note that while this result implies that rather large SFR e-folding times may be common among our SFE E galaxies, significant portions of the distributions in Fig.\ \ref{bcheck} for smaller values of $\tau$ occupy the same region as the distribution for $\tau = \infty$.  In particular, the area of overlap between the $\tau = \infty$ distribution and that for each of the other values of $\tau$ corresponds to 59\%, 87\%, and 96\% of the total area for $\tau =$100 Myr, 300 Myr, and 1 Gyr, respectively.  This implies that while our fitting results are generally consistent with our SFE E galaxies having a tendency for large values of $\tau$, we cannot rule out shorter e-folding times for many of the galaxies.\par

To test the assumption that the limits placed on the fitted parameters lessened or eliminated the possible degeneracies between the age of the younger population, $t_{y}$, and other parameters, we have computed the Spearman rank correlation coefficient among all modelled galaxies for $t_{y}$ and each of the other free parameters.  Of particular concern are the amount of dust extinction, $E(B-V)_{gas}$, and the two metallicity indicators, O/H for the younger population and [MgFe]$^{\prime}$ for the older population.  This is because increases in any of these parameters will in general make the spectra redder, which can roughly mimic the effects of making the populations older.  Therefore, if the fitting process has incorrectly inflated the values of these parameters and compensated by artificially lowering the fitted ages, we would expect to see anti-corrlations between $t_{y}$ and these parameters.  There is a small correlation between $t_{y}$ and $E(B-V)_{gas}$ for the model fits with $\tau \leq 1$ Gyr with correlation coefficients of 0.10, 0.26, and 0.15 for $\tau =$100, 300, and 1000 Myr, respectively.  For the constant SFR model fits, there is a negligible anti-correlation with a coefficient of about -0.017.  Both O/H and [MgFe]$^{\prime}$ are somewhat correlated with $t_{y}$ with correlation coefficients for O/H of about 0.3 for $\tau =$100 and 300 Myr and about 0.1 for $\tau =$1 Gyr and $\infty$ and coefficients for [MgFe]$^{\prime}$ of about 0.3 for all star formation histories.  The relatively small magnitudes of these correlations, coupled with the fact that they are positive, implies that the constraints placed on these parameters have done a reasonable job of eliminating problematic degeneracies between $t_{y}$ and metallicity or the amount of dust extinction.\par
A possible correlation between the age of the younger population, $t_{y}$, and the fraction of the stellar mass attributed to the younger population, $f_{y}$, is a special concern since since such a correlation can exist for both physical and artificial reasons.  For instance, one would naively expect galaxies that have been forming stars longer to have larger values of both $t_{y}$ and $f_{y}$, on average.  On the other hand, as the newly formed population ages, its mass-to-light ratio will increase, implying that if the age of the population is over or under estimated, the fitting routine may be able to compensate by increasing or decreasing $f_{y}$, introducing an artificial correlation.  To explore this, we have used the covariance matrix for the seven free parameters computed by {\it mpfit} to compute the correlation coefficient for $t_{y}$ and $f_{y}$ for each spectral fit.  The mean coefficient is about 0.7 for all star formation histories, indicating a significant but moderate correlation between these parameters which may be mostly artificial.  The effect of this correlation on the model results is discussed below.

\subsection{Model results}
\begin{figure}
\includegraphics[scale=0.42]{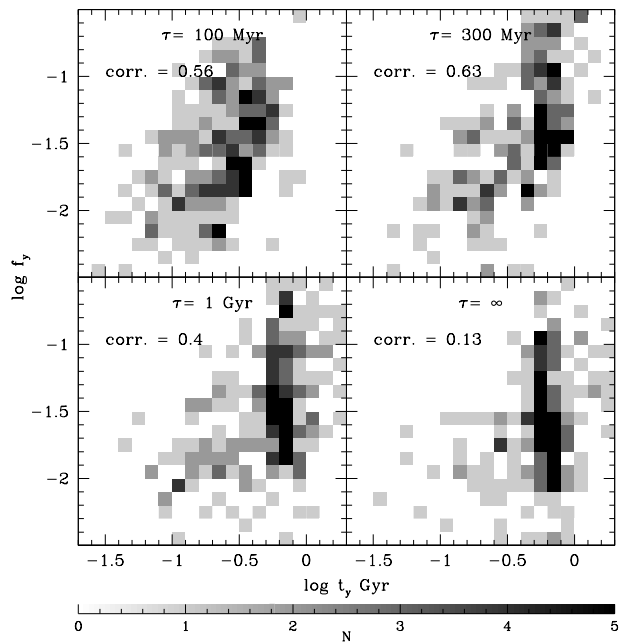}
\caption{The two-dimensional histograms for $t_{y}$ and $f_{y}$ (see \S 4.4) for the SFE E galaxies for each SFR e-folding time, $\tau$.  The Spearman rank correlation coefficient for $t_{y}$ and $f_{y}$ for each value of $\tau$ is printed in each panel as well.}
\label{tyfy}
\end{figure}

The main focus of the spectral fitting procedure detailed above was to estimate the distribution of ages for the younger stellar populations within the SFE E galaxies.  Since the possible artificial correlation between $t_{y}$ and $f_{y}$ will significantly influence the interpretation of this main result, we have displayed the two-dimensional $t_{y}$,$f_{y}$ distribution among all SFE E galaxies for each value of $\tau$ in Fig.\ \ref{tyfy}.  From these plots, it is evident that the typical age of the younger stellar population increases with increasing SFR e-folding time, $\tau$, as one might expect.  For all star formation histories, there is a paucity of galaxies with $t_{y}>1$ Gyr.  However, inspection of the two-dimensional histograms shows that for larger values of $\tau$, this may be influenced somewhat by the $t_{y}$,$f_{y}$ correlation discussed above.  In particular, those galaxies with $\mbox{log } t_{y}/1 \mbox{ Gyr} \approx -0.3$ which have $f_{y}$ values noticeably larger than the location of the peak of the distribution ($\mbox{log } f_{y} \:^{>}_{\sim} -1.3$) may in reality have larger values of $t_{y}$, but the fitting routine artificially increased $f_{y}$ instead of appropriately increasing $t_{y}$ to fit their stellar continua.  We note, however, that only about 8-10\% of the galaxies have $-0.4 < \mbox{log } t_{y}/1 \mbox{ Gyr} < -0.2$ and $\mbox{log } f_{y} > -1.3$ for $\tau \geq 1$ Gyr (see Fig.\ \ref{tyfy}).  This implies that while there is legitimate reason to doubt the estimates for $t_{y}$ for a relatively low number of individual galaxies, the dearth of galaxies with $t_{y}>1$ Gyr appears to be a valid result.\par

\begin{figure}
\includegraphics[scale=0.42]{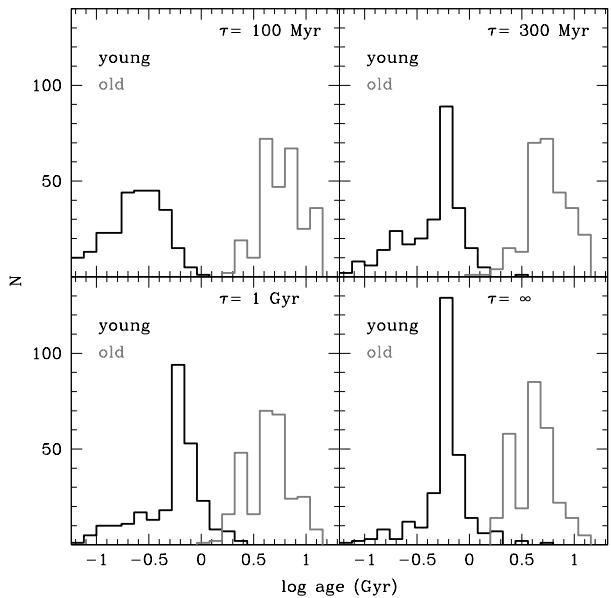}
\caption{The distributions for the ages of the younger (black line) and older (grey line) stellar population derived from the model fits to the stellar continua (see \S 4) for each value of the SFR e-folding time, $\tau$.}
\label{ages}
\end{figure}

To demonstrate that our fitting routine has indeed successfully separated the stellar continua of our SFE E galaxies into older and younger components, we have displayed the distributions for the ages of both the younger and older populations in Fig.\ \ref{ages} for each value of $\tau$.  For all values of $\tau$, the two distributions are noticeably distinct with the difference in ages for the two populations being about 4-5 Gyr, on average.  For $\tau \geq 1$ Gyr, the fraction of galaxies with older population ages $<2$ Gyr, 3-5\%, is larger than the $\leq 1$\% of galaxies with older population ages $<2$ Gyr found for $\tau \leq 300$ Myr.  This is likely an indirect result of the consequences of the $t_{y}$,$f_{y}$ correlation discussed above as the model may force the older populations in some galaxies to be younger to compensate for underestimating $t_{y}$.  However, as noted above, this is only relevant for a relatively low number of SFE E galaxies and does not invalidate the conclusion that there are relatively few SFE E galaxies with younger populations older than 1 Gyr. This main result implies that the star formation within these galaxies typically lasts less than $\sim$1 Gyr.\par

To demonstrate that their relatively short star formation time scales will assure that the SFE E galaxies will become K+A galaxies, we have constructed model spectra using the \citet{bru03} models for a 1 Gyr episode of star formation at a metallicity of Z=0.05 for all values of $\tau$.  We have chosen 1 Gyr because while the typical younger stellar population age is $\sim$600 Myr, the relatively sharp drop in the distribution of ages at $\sim$1 Gyr (see Fig.\ \ref{tyfy} and \ref{ages}) indicates that this is a better choice for the typical star formation episode duration.  The $\tau = 100$ Myr models have SFRs that decrease quickly enough that they already have $W(H\alpha) < 1.6 \mbox{ \AA}$ for ages $> 1$ Gyr regardless of the magnitude of the star formation episode (see below).  This implies that no 1 Gyr SFR cut-off is needed for these models to be able to explain the lack of SFE E galaxies with $t_{y}>1$ Gyr (see Fig.\ \ref{wha}), and therefore, no such cut-off was imposed.  For the other values of $\tau$, such a cut-off is required and was produced by forcing the model SFRs to abruptly go to zero at a time step of 1 Gyr.\par
\begin{figure*}
\includegraphics[scale=0.75]{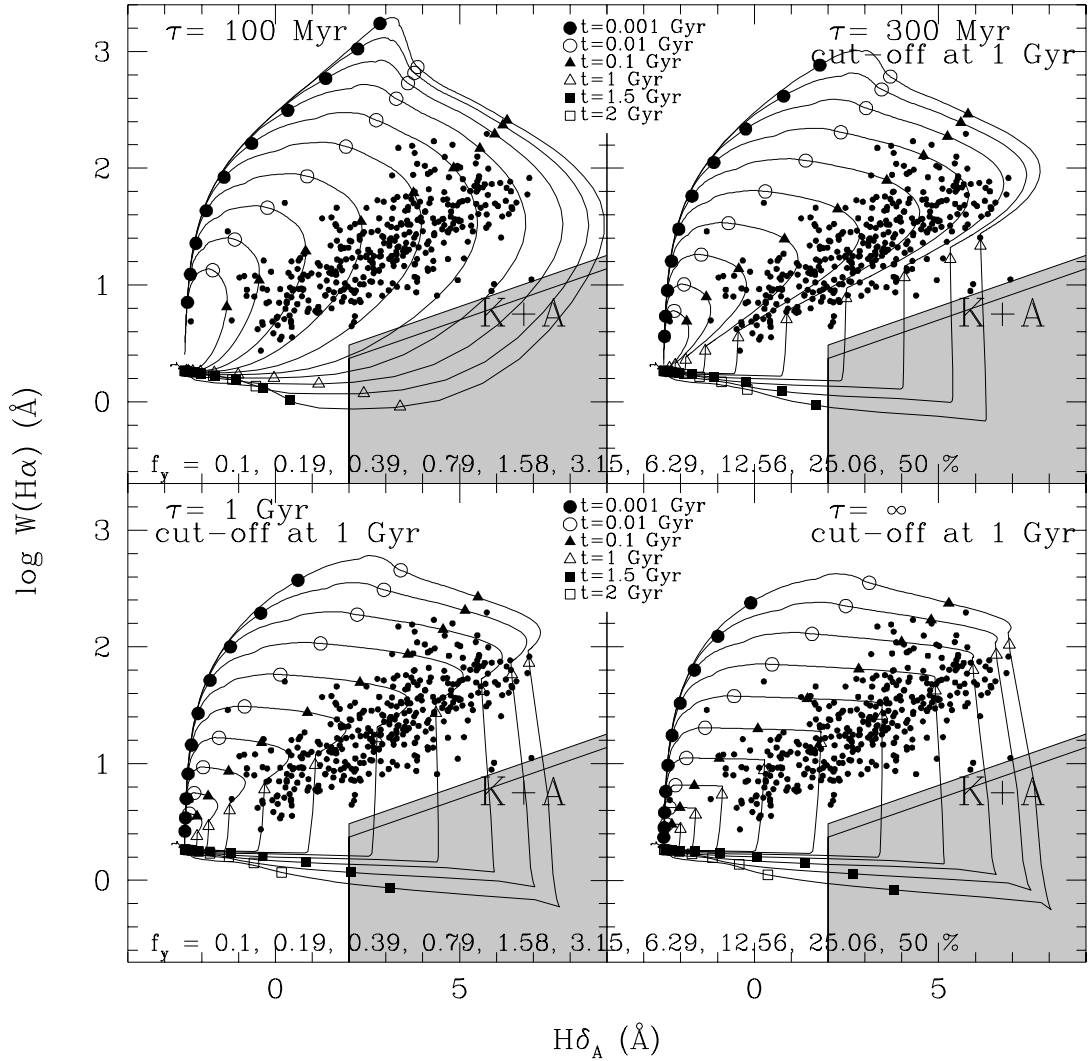}
\caption{For the SFE E galaxies, we plot $W(H\alpha)$ versus $H\delta_{A}$.  Here, $W(H\alpha)$ was corrected for extinction according to log $W(H\alpha)/W(H\alpha)_{obs}=0.4E(B-V)$ \citep{cal01} where the colour excess was derived from the Balmer decrement (see \S 4.2).  Also plotted are model tracks for a 1 Gyr episode of star formation added to a 5 Gyr old stellar population for different values of $f_{y}$ at 1 Gyr (i.e., the total, fractional increase of the stellar mass caused by the star formation episode; see \S 4.3) for each SFR e-folding time, $\tau$.  For $\tau \geq 300$ Myr, the SFR was forced to instantaneously go to zero after 1 Gyr (see \S 4.4).  Times step at 1 Myr, 10 Myr, 100 Myr, 1 Gyr, 1.5 Gyr, and 2 Gyr are highlight with different point styles along each track.  Our definition of K+A galaxies is indicated by the shaded region with two upper limits, the original upper limit given in Fig.\ \ref{classdef}, and the original limit with an extinction correction applied according to \citet{cal01} using the median $E(B-V)$ derived from the Balmer decrement for the SFE E galaxies (see \S 4.2).}
\label{sfhfspec}
\end{figure*}

We have combined these model spectra with the spectrum of a 5 Gyr (the median age of the older populations from the model fits) instantaneous burst population with Z=0.05 for episodes of several different strengths which we characterise by the value of $f_{y}$ at 1 Gyr (i.e., the total fractional increase of the stellar mass).  We have measured both $W(H\alpha)$ and $H\delta_{A}$ for these composite spectra, assuming $T_{e}=5500$ K and case B recombination to convert the values of $Q(H^{0})$ into H$\alpha$ luminosities.  The paths galaxies would take in the $H\delta_{A}$,$W(H\alpha)$ plane according to these model values are shown in Fig.\ \ref{sfhfspec} along with the measured data for the SFE E galaxies for each value of $\tau$.  Each model track starts at $H\delta_{A} = -2.5 \mbox{ \AA}$ and $W(H\alpha) = 1.8 \mbox{ \AA}$, and loops clockwise around the $H\delta_{A}$,$W(H\alpha)$ plane, with larger loops corresponding to larger values of $f_{y}$.  These model tracks show that only those galaxies whose stellar mass will be increased by about 2-5\% or more, depending on the value of $\tau$, will become K+A galaxies according to our definition (i.e., the model tracks will pass through the shaded area in Fig.\ \ref{sfhfspec}).  According to the fitted model spectra, the median fractional increase in the stellar mass for our SFE E galaxies for 1 Gyr of star formation is $\sim4$\% with about 78\% having a fractional increase of more than 2\%.  This is consistent with what has been found by similar, complementary studies conducted with early-type, SDSS galaxies \citep{kav07,nol07,sch07}.  It should be noted that this only applies to the stellar mass within the SDSS fibre aperture since we do not have constraints on the full spatial extent of the star forming regions within these galaxies.  The model tracks also show that the K+A phase ends significantly less than 1 Gyr after the end of the star formation episode, depending of the strength of the episode and the SFR e-folding time.  This is unexpected given the fact that our SFE E and K+A E galaxies are found in roughly the same numbers.  However, we note that for this exercise, we have assumed that the SFR instantaneously goes to zero for $\tau \geq 300$ Myr after 1 Gyr of star formation, when in reality, the SFR may decrease more gradually.  This implies that the time a galaxy will spend as a K+A galaxy estimated from the panels in Fig.\ \ref{sfhfspec} is a lower limit.  We also note that the strict emission line ratio (see \S 2.1 and Fig.\ \ref{diag}) and morphological (see \S 2.2) selection criteria used to identify SFE E galaxies may have caused a significant underestimation of the number of E and S0 galaxies with recent star formation.  This implies that this result does not necessarily contradict the numbers SFE E and K+A E identified for this study (see Table \ref{galnum}).

\section{Star formation time scale and the possible formation of nuclear stellar disks}
On average, the star formation episodes within our SFE E galaxies will only increase their stellar masses by roughly 4\%.  The SFE E galaxies have a median stellar mass within the area of the SDSS 3 arcsec fibre aperture of about $6 \times 10^{9}$ M$_{\odot}$, implying that once the star formation has ceased within these galaxies, the mass of newly formed stars will be about 2$\times 10^{8}$ M$_{\odot}$.  For reference, for the SFE E galaxies, the radius of the SDSS fibre aperture ranges from $0.1R_{e}$ to $2R_{e}$, where $R_{e}$ is the half-light radius from the \citet{bla03b} Sersic fits, or 0.2 to 3 kpc with median values of $0.4R_{e}$ and 1.5 kpc.  While such a relatively small increase in mass will do little to change the galaxies' overall appearances, the approximate stellar mass of newly formed stars is very similar to what has been found for nuclear stellar disks \citep{sco98,piz02}.  The discovery of nuclear stellar disks in the process of being formed would be extremely useful as it is not known whether such disks formed with their host galaxies or whether they formed out of gas deposited at a later epoch.  Many early-type galaxies have been found to contain central gas and/or dust disks \citep{kor89}.  However, only two galaxies, NGC 5845 \citep{kor94} and NGC 4468A \citep{kor05}, that have nuclear stellar and dust disks are known.  There is some indication from the colours of these two disks that they are younger than the surrounding stellar population and that they may have formed more recently.  However, if the relatively weak star formation within our sub-sample of SFE E galaxies can be shown to be confined to compact nuclear disks, this would provide concrete evidence that similar disks typically formed more recently than the rest of the stars within their host galaxies.\par
\begin{figure}
\includegraphics[scale=0.42]{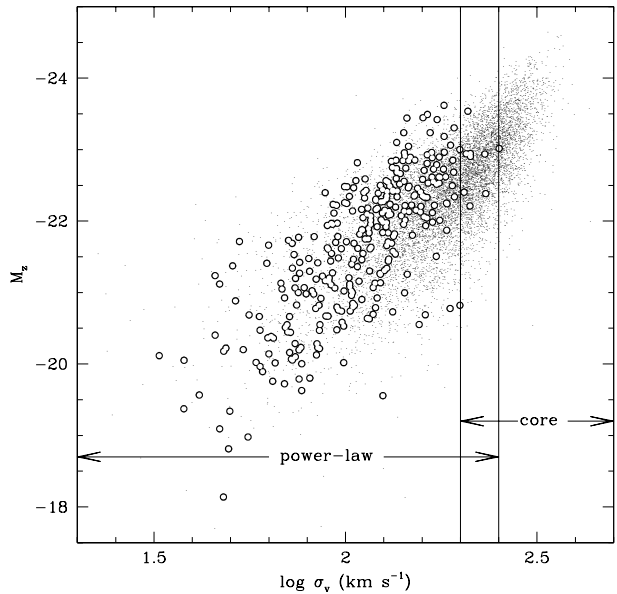}
\caption{The absolute $z$-band magnitude versus the stellar velocity dispersion for all early-type galaxies (black points) and for SFE E galaxies (white circles).  The ranges in velocity dispersion inhabited by power-law and core early-type galaxies from \citet{fab97} are marked for reference.}
\label{amzsv}
\end{figure}

There is other evidence that the SFE E galaxies may harbour nuclear stellar disks that are currently forming.  Central stellar disks are preferentially found within lower-mass early-type galaxies that have steep or "power-law" surface brightness profiles within radii of less than 100 pc \citep{lau95}.  This is in contrast to higher-mass early-type galaxies that have nearly flat or "core" inner profiles \citep{fab97}.  In Fig.\ \ref{amzsv}, we have plotted $z$-band absolute magnitude versus velocity dispersion for all early-type galaxies with $m_{r}<$16 and for our SFE E galaxies.  For reference, we have also highlighted the ranges in velocity dispersion inhabited by power-law and core galaxies taken from \citet{fab97}.  From this plot, it can be seen that there is relatively little overlap between core and power-law galaxies and that the SFE E galaxies lie entirely within the range inhabited by power-law galaxies which tend to be the hosts of nuclear stellar disks.\par
\begin{figure}
\includegraphics[scale=0.42]{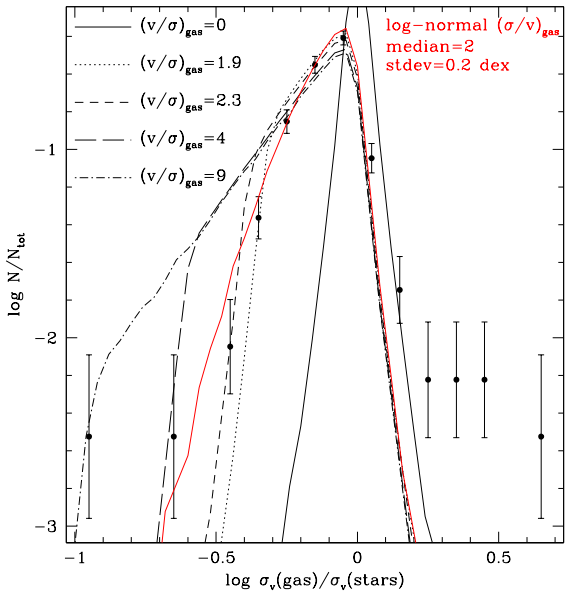}
\caption{The distribution of the ratio of the LOS velocity dispersion of the ionised gas to that of the stars for the SFE E galaxies.  The curves represent model predictions computed using equation (2) for different values of $v_{0} / \sigma_{0}$ for the gas, where $v_{0}$ and $\sigma_{0}$ are the rotational and random velocity components, respectively (see \S 5).  The red curve represents a model that assumes a log-normal distribution for $v_{0} / \sigma_{0}$ with a median of two and a standard deviation of 0.2 dex.}
\label{rot}
\end{figure}

The gas kinematics of the SFE E galaxies are also consistent with the notion that the current star formation in these galaxies is occurring within disk-like structures.  If the star formation is taking place within disks, they are likely rotationally supported.  If this is the case, the effects of inclination will contribute significantly to the shape of the distribution of the ratio of the observed velocity width of the nebular emission lines, $\sigma_{v}$(gas), and the LOS velocity dispersion of the stars, $\sigma_{v}$(stars).  This is because the measured stellar velocity dispersion is likely dominated by the older stellar population since, according to the model results, it typically makes up about 98\% of the total stellar mass and about 70\% of the total V-band luminosity of each SFE E galaxy, while the emission line gas is likely only associated with the younger stellar population.  Because of this, if the younger population is forming within a rotating disk, galaxies that have disks with smaller inclination angles will skew the distribution of $\sigma_{v}$(gas)/$\sigma_{v}$(stars) toward lower values.  In Fig.\ \ref{rot}, we plot the distribution of $\sigma_{v}$(gas)/$\sigma_{v}$(stars) for the SFE E galaxies, where values for $\sigma_{v}$(gas) were taken from \citet{tre04} and the values of $\sigma_{v}$(stars) were measured using {\it vdispfit} (see \S 3).  From this figure, it can be seen that the distribution is indeed asymmetric with a larger fraction of galaxies having  $\sigma_{v}$(gas)/$\sigma_{v}$(stars)$<$1.\par
To explore the possibility that this asymmetry is indicative of rotating gas, we have constructed the following simple models.  We first made the simplifying assumption that the stellar distribution is "pressure" supported, i.e., $\sigma_{v}(\mbox{stars})=v_{c}$, where $v_{c}$ is the circular velocity.  We then allowed the gas to have both random and ordered (i.e., rotational) velocity components, which are related to the circular velocity according to $v_{c}^{2}=\sigma_{0}^{2}+v_{0}^{2}$ where $\sigma_{0}$ and $v_{0}$ are the random and ordered components, respectively.  In this case, the LOS velocity dispersion of the gas is given by $\sigma_{v}$(gas)=$\sqrt{\sigma_{0}^{2}+v_{0}^{2} \mbox{sin}(i)^{2}}$ where $i$ is the inclination angle for the axis of rotation of the gas.  Combining the above relations yields the following\par
\begin{equation}
\left ( \frac{\sigma_{v} (\mbox{gas})}{\sigma_{v} (\mbox{stars})} \right ) ^{2} = \frac{\left [ \frac{v_{\circ}}{\sigma_{\circ}} \mbox{sin}(i) \right ]^{2} + 1}{\left ( \frac{v_{\circ}}{\sigma_{\circ}} \right )^{2} + 1}
\label{vrateq}
\end{equation}

Using equation (\ref{vrateq}), we have constructed several model distributions for $\sigma_{v}$(gas)/$\sigma_{v}$(stars) for different values for $v_{\circ}/\sigma_{\circ}$.  For all models, we assumed that $i$ was distributed according to sin$(i)$, the distribution one would expect for random orientations \citep{hub26}.  To incorporate the effects of observational uncertainty, we fit a Gaussian to the distribution of errors in $\sigma_{v}$(gas)/$\sigma_{v}$(stars) for the SFE E galaxies and convolved each model distribution with this Gaussian.  These model distributions are plotted with the data in Fig.\ \ref{rot} with the Gaussian kernel plotted as a model where $v_{\circ}/\sigma_{\circ}=0$.  The figure demonstrates that in order to reproduce the observed distribution, some amount of rotation is required for the gas.  The model curves show that a single value of $v_{\circ}/\sigma_{\circ}$ cannot reproduce the entire observed distribution and that values ranging from about two to ten are required.  To better illustrate this, we have also included another model curve (the red curve in Fig.\ \ref{rot}) which used a log-normal distribution of values for $v_{\circ}/\sigma_{\circ}$ with a median of two and a dispersion of 0.2 dex.  This model curve is a close approximation to the observed distribution for $0.2 < \sigma_{v}(\mbox{gas})/\sigma_{v}(\mbox{stars}) < 1.3$ and implies that $v_{\circ}/\sigma_{\circ}$ ranges from 0.8 to five with a typical value near two.\par
We also constructed model distributions for a scenario where the stars also have a rotational component and $v_{\circ}/\sigma_{\circ}$ for the gas has a log-normal distribution.  For this exercise, we fixed $v_{\circ}/\sigma_{\circ}$ for the stars at 0.65, the amount of rotation needed to produce the maximum allowed ellipticity of 0.3 (see \S 2.2).  To reproduce the red curve in Fig.\ \ref{rot} with this model, it was necessary to increase the median value of $v_{\circ}/\sigma_{\circ}$ for the gas to 2.3-2.5, depending on how the rotation axes of the stars and gas were aligned.  This implies that the estimated typical value of $v_{\circ}/\sigma_{\circ} \approx 2$ for the gas is a lower limit.  This further demonstrates that not only does the observed distribution indicate the presence of rotation within the gas, it also implies that the rotational component of the gas velocity is the dominant component.  We note that the observed distribution does have a tail for $\sigma_{v}$(gas)/$\sigma_{v}$(stars)$>$1.6 that is not reproduced by any of the models.  However, this tail consists of only 2\% (7 galaxies) of the SFE E sub-sample, all but one of which have errors in their gas velocities that are unusually large (between 15\% and 100\%).  We therefore conclude that the data are consistent with the star forming regions in SFE E galaxies typically being contained within rotationally supported disks.\par
Since our SFE E galaxies have no prominent disks visible in their $g$-band images, it is reasonable to infer that the disks inferred from Fig.\ \ref{rot} may be relatively small and similar to known nuclear stellar and gas disks.  In fact, the estimated typical value of $v_{\circ}/\sigma_{\circ} \approx 2$ for the gas is consistent with what has been observed for the stellar populations in the centres (radii $<100$ pc) of early-type galaxies with identified nuclear stellar disks where $v_{\circ}/\sigma_{\circ} \sim 1$ \citep[see, e.g.,][]{kra04}.  However, the fact that the observed distribution of $\sigma_{v}(\mbox{gas})/\sigma_{v}(\mbox{stars})$ peaks near unity implies that the ionised gas probes the same range in circular velocity as the stars.  Since the SDSS fibre aperture extends to radii of at least 200 pc and up to 3 kpc (or 0.1 to two effective radii) for our SFE E galaxies, that result seems to imply that the disks may be somewhat extended.  But, we must also recognise the fact that $\sigma_{v}$(gas) and $\sigma_{v}$(stars) are both luminosity weighted velocity dispersions.  Since the mean Sersic index for our SFE E galaxies is about 4.5, the stellar velocity dispersions must be biassed toward smaller radii, implying that the H$\alpha$ surface brightness profiles must be similarly biassed.  Therefore, even though some spatially extended star formation may exist within these galaxies, they likely have some centrally concentrated star formation.  This makes our SFE E galaxies excellent candidates for follow-up narrow-band imaging and spectroscopy to search for nuclear disks with active star formation.

\section{Discussion and conclusions}
Using the vast amount of galaxy imaging and spectroscopic data available from the SDSS, we have selected a relatively large, clean sample of actively star-forming E and S0 galaxies.  These galaxies make up about 3\% of all early-type SDSS galaxies with $m_{r}<$16 and 0.7\% of all Main \citep{str02} galaxies with $m_{r}<$16.  These galaxies appear to be less massive on average than typical early-type galaxies which is consistent with what has been found by the similar, complementary study of \citet{sch07}.  This is also consistent with what has been found by other authors \citep[e.g.,][]{tre05,gra07}, that the fraction of stellar mass formed at low redshift is substantially higher for lower mass early-type galaxies.  The number of star forming E and S0 galaxies is also comparable with the number of morphologically similar K+A galaxies.  The relatively complex selection criteria used to select both star forming E and S0 galaxies and K+A galaxies (see \S 2) make it difficult to compare the sample sizes in a statistically robust way.  However, the fact that the numbers of both types of galaxies are not vastly different does allow for the possibility that the star forming E and S0 galaxies are the progenitors of E and S0 K+A galaxies.\par
Through modelling of the stellar continua of our SFE E galaxies, we have demonstrated that this is indeed plausible.  We have shown that our star-forming E and S0 galaxies likely form stars for$\;^{<}_{\sim}$1 Gyr.  The model results demonstrate that the factor that most strongly determines whether or not one of these galaxies will become a K+A galaxy is the magnitude of its star formation episode.  According to our K+A definition, only star formation episodes that increase the stellar mass by roughly 2-5\% or more will lead to K+A galaxies, which appears to be the case for about 80\% of the SFE E galaxies.  This suggests that in terms of star formation episode duration, all star-forming E and S0 galaxies are similar to E and S0 K+A galaxies, even though some of them will not have Balmer absorption lines that are strong enough for them to be classified as K+A galaxies after their star formation has ceased.\par
With the implication that our samples of star-forming and K+A E and S0 galaxies are linked in an evolutionary sequence comes a large potential for followup observations.   For instance, as discussed in \S 5, there is evidence to suggest that star-forming E and S0 galaxies may be the sites of nuclear stellar disk formation.  High spatial resolution narrow-band images of the star-forming E and S0 galaxies will be able to determine whether or not this is true.  If so, E and S0 K+A galaxies will be excellent candidate hosts of nuclear stellar disks.  If such disks are responsible for the galaxies' strong Balmer absorption, then they should appear relatively young (i.e., bluer than the surrounding stars); this may also be explored will high spatial resolution followup imaging.\par
Identifying relatively large samples of SFE E galaxies with these types of disks will prove extremely useful for measuring central black hole masses with future spectroscopic observations (e.g., using NIRSpec with the James Webb Space Telescope) for relatively low mass E and S0 galaxies.  This is because the kinematics of these nuclear disk structures are dominated by rotation and velocity anisotropy is relatively unimportant \citep{kor05}.  However, estimates for BH masses using nuclear stellar disks are based on luminosity weighted stellar velocities measured from optical spectra which are affected by radial variations in the stellar population \citep{kor95}.  Rotation velocities measured for gas using emission lines are not affected by this phenomenon.  In this sense, the interpretation of the kinematics of any emission line gas associated with a nuclear disk in regards to the mass of the central BH will be much more straightforward.  Identifying SFE E galaxies that contain such disks will therefore provide a sample for which future spectroscopy can be used to reliably estimate central BH masses.

\section*{Acknowledgements}
The authors would like to thank the referee for detailed comments that improved the paper.  This research was partially performed while the lead author held a National Research Council Research Associateship Award at the Naval Research Laboratory.  Basic research in astronomy at the Naval Research Laboratory is supported by 6.1 base funding.\par
Funding for the creation and distribution of the SDSS Archive has been provided by the Alfred P. Sloan Foundation, the Participating Institutions, the National Aeronautics and Space Administration, the National Science Foundation, the U.S. Department of Energy, the Japanese Monbukagakusho, and the Max Planck Society. The SDSS Web site is http://www.sdss.org/.\par
The SDSS is managed by the Astrophysical Research Consortium (ARC) for the Participating Institutions. The Participating Institutions are The University of Chicago, Fermilab, the Institute for Advanced Study, the Japan Participation Group, The Johns Hopkins University, the Korean Scientist Group, Los Alamos National Laboratory, the Max-Planck-Institute for Astronomy (MPIA), the Max-Planck-Institute for Astrophysics (MPA), New Mexico State University, University of Pittsburgh, University of Portsmouth, Princeton University, the United States Naval Observatory, and the University of Washington.

\bsp

\label{lastpage}

\end{document}